\documentclass{article}
\usepackage[margin=3cm]{geometry}

\usepackage{amsmath,amssymb,amsfonts}
\usepackage{physics} 

\usepackage{makecell}
\setcellgapes{1ex}
\usepackage{cases}

\usepackage{authblk}

\usepackage{graphicx,color}
\usepackage{epstopdf}

\usepackage{float}

\usepackage[nocompress]{cite}

\usepackage[shortlabels]{enumitem}

\newcommand{\ourG}{\mathbf{G}}
\newcommand{\ourB}{\mathbf{B}}
\newcommand{\LL}{\mathcal{L}}

\numberwithin{equation}{section}

\usepackage{hyperref}
\hypersetup{colorlinks=true,allcolors=blue,urlcolor=cyan}

\title{Cosmological fluids with boundary term couplings}

\author[1,2]{Christian G B\"ohmer\footnote{Email: c.boehmer@ucl.ac.uk}}
\author[1]{Antonio d'Alfonso del Sordo\footnote{Email: a.dalfonsodelsordo@ucl.ac.uk}}

\affil[1]{Department of Mathematics, University College London, \authorcr 
Gower Street, London WC1E 6BT, UK\medskip}
\affil[2]{Astrophysics Research Centre, School of Mathematics, \authorcr 
Statistics and Computer Science, University of KwaZulu-Natal, \authorcr 
Private Bag X54001, Durban 4000, South Africa\medskip}

\date{1 June 2024} 

\pagecolor{white}

\begin{document}

\renewcommand{\arraystretch}{1.2} 
\setlength{\tabcolsep}{1ex} 
\setlength{\extrarowheight}{1ex} 

\maketitle

\begin{abstract}
Cosmological models can be studied effectively using dynamical systems techniques. Starting from Brown's formulation of the variational principle for relativistic fluids, we introduce new types of couplings involving a perfect fluid, a scalar field, and boundary terms. We describe three different coupling models, one of which turns out to be particularly relevant for cosmology. Its behaviour is similar to that of models in which dark matter decays into dark energy. In particular, for a constant coupling, the model mimics well-known dynamical dark energy models while the non-constant couplings offer a rich dynamical structure, unseen before. We are able to achieve this richness whilst working in a two-dimensional phase space. This is a significant advantage which allows us to provide a clear physical interpretation of the key features and draw analogies with previously studied models. 
\end{abstract}

\clearpage 

\section{Introduction}

Cosmology, the scientific study of the universe as a whole, has undergone remarkable advances in recent decades and General Relativity (GR) provides a good model to describe cosmological gravitational phenomena~\cite{LIGOScientific:2016aoc,Planck:2018vyg,Will:2018bme,Ishak:2018his}.
On the other hand, open questions in cosmology remain, foremost among which are the dark energy and dark matter problems. The nature of dark energy, which is responsible for driving the universe's late-time accelerated expansion, is not well understood, and it often assumed to be a cosmological constant. Since the first observational evidence of an accelerated expansion~\cite{SupernovaSearchTeam:1998fmf,SupernovaCosmologyProject:1998vns} of the universe, a plethora of cosmological models to explain dark energy has emerged, for a review see~\cite{Copeland:2006wr}. 

The addition of a positive cosmological constant $\Lambda$ to the Einstein field equations, originally introduced by Einstein~\cite{Einstein:1917ce} for his static universe, is one of the most straightforward candidates for dark energy. This paves the way for the $\Lambda$ Cold Dark Matter ($\Lambda$CDM) model.  However, the $\Lambda$CDM model fails to explain why the inferred value of $\Lambda$ is so small compared to the vacuum energy density expected from particle physics~\cite{Weinberg:1988cp}. It is also unclear why its value is comparable to the matter density today. This constitutes the so-called \emph{coincidence problem}~\cite{Zlatev:1998tr,Sadjadi:2006qp}. 

One way to begin to address this issue is to allow for a dynamical cosmological constant~\cite{Copeland:2006wr}, that is, to introduce some dynamical field able to reproduce the late-time acceleration behaviour and mimic the properties of the cosmological constant. The simplest such model is a canonical scalar field $\phi$ with flat potential $V(\phi)$, which drives the accelerated expansion of the universe. Any model of this type is referred to as \textit{quintessence}~\cite{Tsujikawa:2013fta}. Scalar fields play a major role in modern cosmology as they are also able to drive inflation, the early-time epoch of accelerated expansion~\cite{Tamanini:2014mpa,Bohmer:2016ome,Urena-Lopez:2020npg}. Scalar field models have also been used as candidates for dark matter models, see~\cite{Magana:2012ph}. We are primarily interested in scalar fields as models to drive a period of accelerated expansion, at both early and late times of the universe's evolution.

Another approach is to consider dark energy as evidence for the incompleteness of GR and, hence, seek extensions or modifications of GR~\cite{CANTATA:2021ktz,Koyama:2018som,Joyce:2014kja}. Several models to describe the dark energy interaction with dark matter have been proposed~\cite{Billyard:2000bh,Farrar:2003uw,Guo:2007zk,Caldera-Cabral:2008yyo,He:2008tn,Pereira:2008at,Valiviita:2009nu,Bamba:2012cp} while some authors, e.g.~\cite{Joyce:2016}, have emphasised a strong distinction between modified theories of gravity and dark energy models. 

From a theoretical point of view, in GR, one usually restricts the Lagrangian to a linear function of the Ricci scalar, minimally coupled with matter. However, there is no reason, \textit{a priori}, to assume such a restriction. So, one can modify the gravitational part of the action to allow non-linear corrections to the Lagrangian~\cite{Magnano:1987zz,Capozziello:2007ec}, this is the general approach followed by $f(R)$-theories of gravity~\cite{Sotiriou:2008rp,DeFelice:2010aj}. Some other extensions of GR increase the number of spacetime dimensions or introduce non-minimal matter couplings to boundary and topological terms~\cite{Capozziello:2002rd,Ferraro:2006jd,Nojiri:2006je,Nojiri:2010wj,Harko:2011kv,Clifton:2011jh,Granda:2014zea,Bahamonde:2015hza,Nojiri:2017ncd,Boehmer:2021aji}. These are terms in the Lagrangian that describe how matter couples to geometrical quantities. Non-minimally coupled terms involving curvature vanish in the limit of special relativity. 

We will follow the approach suggested in~\cite{Boehmer:2015kta,Boehmer:2015sha}, and, therefore, build models of quintessence interacting with dark matter. This involves introducing couplings at the level of the action which characterise both quintessence and dark matter~\cite{Boehmer:2008av,Tamanini:2015iia}. In particular, we extend the Lagrangian formulation by Brown~\cite{Brown:1992kc} which describes a perfect fluid~\cite{Koivisto:2015qua}. Motivated by~\cite{Boehmer:2015kta,Boehmer:2015sha}, we introduce new couplings containing a boundary term and a pseudovector related to the boundary term. 

In~\cite{Bahamonde:2015hza,Kadam:2023ufk}, a dynamical system analysis where teleparallel quintessence is non-minimally coupled to a boundary term is presented. In the same spirit, we study the background cosmology within this framework and apply dynamical systems tools to investigate the dynamics of the different models. Our ultimate goal is to examine the behaviour of these dark energy models.  Dynamical systems theory has emerged as a vital tool in cosmology~\cite{Bahamonde:2017ize}, and has been employed successfully to study modified theories of gravity in the cosmological context ~\cite{Boehmer:2022wln,Dutta:2017fjw,Khyllep:2021pcu}. 

The paper is organised as follows. In Section~\ref{sec:lagrangian}, we present the Lagrangian formulation of our models, and, in Section~\ref{sec:cosmology}, we focus on the cosmological field equations and introduce the cosmological variables which will be used for our analysis. Section~\ref{sec:constantint} contains our analysis of the dynamical systems for a constant interaction term, and we highlight the analogies with previous models~\cite{Copeland:1997et}. In Section~\ref{sec:nonconstint}, we present the rich and novel dynamical structure in the case of a non-constant interaction term, which, for some choice of the parameters, features both early-time and late-time accelerated expansion. In Section~\ref{sec:disc}, we discuss our results and suggest potential directions for future investigations. 

\paragraph{Notation and conventions.}

Unless otherwise specified, we employ standard relativistic notation throughout.  The signature of the metric tensor $g_{\mu\nu}$ is assumed to be $(-,+,+,+)$, Greek indices are space-time indices taking values in $\{0, 1, 2, 3\}$. The coupling constant appearing in the Einstein field equations is denoted by $\kappa=8\pi G/c^4$, where $c$ is the speed of light and $G$ the Newton's gravitational constant. We use natural units, with $c=1$ and $G=1$. A dot denotes differentiation with respect to cosmological time, a prime denotes the derivative with respect to the argument, or in the case of the dynamical system equations a prime denotes a derivative with respect to the logarithm of the scale factor $\log(a)$.

\section{Lagrangian formulation and field equations}
\label{sec:lagrangian}

\subsection{Gravitational and fluid action}

It is well known that a total derivative term can be isolated from the Ricci scalar, yielding the \emph{Gamma squared action}. This action also gives rise to the Einstein field equations when variations with respect to the metric are considered. However, the underlying Lagrangian is no longer a coordinate scalar as it differs from a true scalar by the total derivative term. We prepend the word `pseudo' to highlight quantities which appear to be scalars or tensors but are not invariant under general coordinate transformations. As shown in \cite{Boehmer:2021aji}, this allows one to write the Einstein-Hilbert action as
\begin{equation}
    S_{\mathrm{EH}} = \frac{1}{2\kappa}
    \int \LL_{\mathrm{EH}}\, \dd^4 x= \frac{1}{2\kappa}
    \int R\sqrt{-g}\, \dd^4 x=
    \frac{1}{2\kappa}\int (\ourG+\ourB)\sqrt{-g}\,\dd^4 x\,.
    \label{eqn:EH}
\end{equation}
The \emph{bulk term} $\ourG$ is defined as
\begin{equation}
    \ourG := g^{\mu\nu}
    \left(\Gamma^\lambda_{\mu\sigma}\Gamma^\sigma_{\lambda\nu}-\Gamma^\sigma_{\mu\nu}\Gamma^\lambda_{\lambda\sigma}\right)\,,
\end{equation}
and the \emph{boundary term} $\ourB$ is given by
\begin{equation}
    \ourB := \frac{1}{\sqrt{-g}}\partial_\nu\left(\frac{\partial_\mu\left(g g^{\mu\nu}\right)}{\sqrt{-g}}\right)=\frac{1}{\sqrt{-g}}\partial_\sigma\left(\sqrt{-g}B^\sigma\right), 
    \label{eqn:defnofBmu}
\end{equation}
where we have introduced the boundary pseudovector $B^\sigma$ given by
\begin{equation}
    B^\sigma = g^{\mu\nu}\Gamma^\sigma_{\mu\nu}-g^{\sigma\nu}\Gamma^\lambda_{\lambda\nu}\,.
    \label{boundaryvector}
\end{equation}
We note that $\ourG$ and $\ourB$ are pseudoscalars.

By construction, the bulk term $\ourG$ is quadratic in the Christoffel symbols and hence the action
\begin{equation}
    S_{\mathrm{E}}=\frac{1}{2\kappa}\int \ourG\sqrt{-g}\,\dd^4 x =
    \frac{1}{2\kappa}\int g^{\mu\nu}
    (\Gamma^\lambda_{\mu\sigma}\Gamma^\sigma_{\lambda\nu}-\Gamma^\sigma_{\mu\nu}\Gamma^\lambda_{\lambda\sigma}) 
    \sqrt{-g}\,\dd^4 x \,,
\end{equation}
is called the Gamma squared action or, sometimes, the Einstein action, to distinguish it from the Einstein-Hilbert action. Recent progress was made in~\cite{Boehmer:2021aji,Boehmer:2023fyl} on constructing modified theories of gravity based on this decomposition. These models can be linked naturally to a variety of other modified gravity models, either within the context of GR or in the metric-affine framework.

The Christoffel symbols are usually interpreted as the gravitational field strengths. We can motivate this by recalling that they contain the first partial derivatives of the metric, which represent the gravitational potentials. The bulk term is thus quadratic in the field strengths, similar to other field theories, like Yang-Mills theories or elasticity theory. This analogy provides the primary motivation for splitting the Ricci scalar as in~(\ref{eqn:EH}). This split naturally yields a boundary term which could be coupled to other fields present in the model. Couplings of this type are interesting as they are purely geometrical and thus have no direct links with classical physics. This is similar to Brans--Dicke theories, where a scalar field is coupled to the curvature scalar. By isolating the bulk and boundary terms, we can therefore consider more intricate couplings involving those two parts, which make up the curvature scalar.

Matter is introduced into the theory by considering a total action of the form \begin{equation}
    S_{\mathrm{tot}} = S_{\mathrm{EH}} + S_{\mathrm{matter}}\,,
\end{equation}
where $S_{\mathrm{matter}}$ is the matter action. This gives rise to the energy-momentum tensor 
\begin{equation}
    T_{\mu \nu} := \frac{-2}{\sqrt{-g}}
    \frac{\delta S_{\mathrm{matter}}}{\delta g^{\mu \nu}}\,.
\end{equation}
Generally, a matter action would depend on matter fields, and variations with respect to those matter fields yield the equations of motion of the matter component. As we will see in the following, this is non-trivial if one wishes to model relativistic fluids using the variational approach. 

Brown~\cite{Brown:1992kc} introduced a Lagrangian formalism for relativistic perfect fluids based on the Lagrangian (density) given by
\begin{equation}
    \LL_{\mathrm{fluid}} = -\sqrt{-g}\rho(n,s) + J^\mu\left(\varphi_{,\mu}+s\theta_{,\mu}+\beta_A\alpha^A_{,\mu}\right), \label{lagrangianmatter}
\end{equation}
where
\begin{itemize}
    \item $n$ is the particle number density
    \item $s$ is entropy density per particle
    \item $\rho(n,s)$ is the energy density of the matter fluid, a function of $n $ and $s$
    \item  $J^\mu$ is the vector-density particle-number flux, which is related to $n$ by
    \begin{equation}
        J^\mu=\sqrt{-g}\,n\,U^\mu\,, \qquad 
        |J|=\sqrt{-g_{\mu\nu}J^\mu J^\nu}\,, \qquad 
        n=|J|/\sqrt{-g}\,, 
        \label{defnofJandn}
    \end{equation}
    where $U^\mu$ is the fluid's $4$-velocity satisfying $U_\mu U^\mu=-1$
    \item  $\varphi$, $\theta$, and $\beta_A$ are all Lagrange multipliers with $A$ taking values in $\{1, 2, 3\}$, and the components $\alpha^A$ are the Lagrangian coordinates of the fluid.
\end{itemize}
The independent dynamical variables of the Lagrangian~\eqref{lagrangianmatter} are $g^{\mu\nu}$, $J^\mu$, $s$, $\varphi$, $\theta$, $\beta_A$, and $\alpha^A$. We note that, in this approach, the pressure of the fluid $p$ is defined as 
\begin{equation}
    p:=n\pdv{\rho}{n}-\rho\,, 
    \label{eq:defofpressure}
\end{equation}
which is consistent with the first law of thermodynamics. 

\subsection{Total action and interaction terms}

We can now set up the total action which contains gravity, a fluid, a scalar field $\phi$, and an interaction term. This means
\begin{equation}
    S_{\mathrm{tot}}= \int ( \LL_{\mathrm{EH}}+\LL_{\mathrm{fluid}}+\LL_{\phi}+\LL_{\mathrm{int}}) 
    \dd^4 x\,, 
    \label{totalaction}
\end{equation}
where $\LL_\phi$ is the scalar field Lagrangian (density) given by
\begin{equation}
    \LL_\phi = -\sqrt{-g}\Bigl(\frac{1}{2}g^{\mu\nu}\nabla_\mu\phi\nabla_\nu\phi+V(\phi)\Bigr), 
    \label{Lphi}
\end{equation}
with given scalar field potential $V$. The Lagrangian (density) $\LL_{\mathrm{int}}$ is an interaction coupling term, which allows us to couple the fluid to the scalar field.

Due to the presence of the various independent variables in Brown's approach, one can propose some types of coupling terms which do not exist in other settings. Moreover, such terms have no natural special relativistic analogue, making this potentially interesting in the context of cosmology. In previous work~\cite{Boehmer:2015kta,Boehmer:2015sha}, one of the authors proposed interaction terms of the form $f(n,s,\phi)$ and $f (n,s,\phi)J^\mu \partial_\mu \phi$. These models gave rise to some unexpected dynamics. In particular, as we wish to take into account boundary terms, we identified the following terms as the suitable possibilities
\begin{enumerate}[(a)]
    \item algebraic scalar coupling: 
    $-\sqrt{-g}f(n,s,\phi,\ourB)$
    \item algebraic vector coupling: 
    $-\sqrt{-g}f(n,s,\phi)B^\mu J_\mu$ 
    \item derivative coupling: 
    $\LL_{\mathrm{int}} := -\sqrt{-g}f(n,s,\phi)B^\mu \partial_\mu \phi$.
\end{enumerate}
Note that $B^\mu$ was defined in~(\ref{boundaryvector}). Depending on the specific interaction term chosen, one should also note that the physical dimensions of $f$ differ in the different couplings. 

Coupling (b) is very restrictive in the context of cosmology. We find that the consistency of the cosmological equations implies that $f$ is proportional to $n$, thereby eliminating the scalar field from the coupling. Consequently, we find equations which largely coincide with the standard cosmological equations, and the model does not exhibit novel behaviour.

For the remainder of this paper, we will consider the interaction term~(c) and $\LL_{\mathrm{int}}$ will denote this interaction Lagrangian. This term was found to have behaviour relevant to cosmology, and gave rise to manageable cosmological equations. In principle, our analysis can be repeated for term (a), and potentially for more complicated terms. For example, $f$ could contain an explicit dependence on $\ourG$, or the Ricci scalar, or there could be higher order couplings containing terms like $(B^\mu J_\mu)(B^\nu \partial_\nu \phi)$, etc. 

\subsection{Variations and field equations}

We begin with the variations of action~(\ref{totalaction}) with respect to the the fields $\varphi$, $\theta$, $\beta_A$, and the Lagrangian coordinates $\alpha^A$, respectively. This yields
\begin{align}
    \delta\varphi:&\quad J^\mu_{,\mu}=0 \,,
    \label{varwrtvarphi}\\
    \delta\theta:&\quad \left(sJ^\mu\right)_{,\mu}=0 \,,
    \label{varwrttheta}\\
    \delta\alpha^A:&\quad \left(\beta_A J^\mu\right)_{,\mu}=0 \,,
    \label{varwrtalphaA}\\
   \delta\beta^A:&\quad \alpha^A_{,\mu} J^\mu=0 \,.
   \label{varwrtbetaA}
\end{align}
These equations are independent of the gravitational action and are also independent of the interaction term. Next, variations of~(\ref{totalaction}) with respect to the entropy density, $s$, give
\begin{equation}
    \delta s:\quad nU^\mu\theta_{,\mu}=\frac{\partial\rho}{\partial s}+\frac{\partial f}{\partial s}B^\mu\partial_\mu\phi \,,
\end{equation}
where the final term depends on the choice of $f$. Variations with respect to $J^\mu$ yield
\begin{equation}
  \delta J^\mu:\quad \varphi_{,\mu}+s\theta_{,\mu}+\beta_A\alpha^A_{,\mu}+\frac{\partial \rho}{\partial n}U_{\mu}+\frac{\partial f}{\partial n}U_{\mu}B^\sigma\partial_{\sigma}\phi=0 \,.
\end{equation}
Again, we have one term which depends on the coupling.

Variations with respect to the scalar field $\phi$ yield a modified Klein Gordon equation
\begin{equation}
     \delta\phi:\quad  \Box \phi-V'(\phi)-\frac{\partial f}{\partial\phi}B^\mu\partial_{\mu}\phi+\nabla_\mu (f B^\mu)=0\,, \label{eq:modifiedKG}
\end{equation}
where $\Box:=\nabla^\mu\nabla_\mu$. 

Finally, variations with respect to the metric tensor yield the Einstein field equations 
\begin{equation}
  \delta g^{\mu\nu}:\quad G_{\mu\nu}=
  \kappa(T^{(\rm fluid)}_{\mu\nu} + T_{\mu\nu}^{(\phi)} + T_{\mu\nu}^{(\mathrm{int})}) \,, 
  \label{eq:EFE}
\end{equation}
where $G_{\mu\nu}$ is the Einstein tensor and 
\begin{align}
    T^{(\rm fluid)}_{\mu\nu} &= 
    (\rho+p)U_{\mu}U_{\nu} + p g_{\mu\nu}, \\
    T_{\mu\nu}^{(\phi)} &= 
    \partial_\mu \phi\partial_\nu \phi -
    g_{\mu\nu}\Bigl(
    \frac{1}{2}\partial_\mu\phi\partial^\mu\phi + V(\phi)
    \Bigr) \,.
\end{align}
Both are the standard forms of the energy-momentum tensors of a perfect fluid and a scalar field, respectively. The energy-momentum tensor related to the interaction term is more complicated and is given by
\begin{multline}
    T^{(\mathrm{int})}_{\mu\nu} = g_{\mu\nu}f(n,s,\phi)B^\sigma\partial_\sigma\phi +\frac{1}{2}n\frac{\partial f}{\partial n}\left(U_\mu U_\nu+g_{\mu\nu}\right)B^\sigma\partial_\sigma \phi \\
    -2\sqrt{-g}\,g_{\mu\nu}g^{\alpha\beta}\partial_\alpha\left(\frac{1}{\sqrt{-g}}f(n,s,\phi)\partial_\beta \phi\right)+2\sqrt{-g}\,\partial_{(\mu}\left(\frac{1}{\sqrt{-g}}f(n,s,\phi)\partial_{\nu)} \phi\right).
    \label{eqn:Tint}
\end{multline}
The second line of this tensor appears due to variations of the boundary pseudovector with respect to the metric. This requires integration by parts and thus leads to the second derivative terms of the scalar field. We note that one could rewrite the partial derivatives of the metric determinants using the Christoffel symbols. However, for the purposes of this work, this would not introduce additional insights.

We motivated the introduction of $\LL_{\mathrm{int}}$ as a new interaction term which could model an interaction between the fluid and the scalar field. However, one can adopt a different interpretation, namely to view~(\ref{eqn:Tint}) as an independent fluid of unusual form.

\section{Cosmological field equations}
\label{sec:cosmology}

In this section, we provide a brief overview of the necessary background material required to study the cosmological field equations of our coupled models. We do this via a dynamical systems formulation, which has proved to be a powerful tool when studying cosmological equations.

In line with current observational evidence \cite{Planck:2018vyg,Saadeh:2016sak,Efstathiou:2020wem}, let us begin with the homogeneous, isotropic, and spatially flat Friedmann-Lema\^itre-Robertson-Walker (FLRW) line element
\begin{equation}
    \dd s^2=-N^2(t)\dd t^2+a^2(t)(\dd x^2+\dd y^2+\dd z^2) \,,
\end{equation}
where $a(t)$ is the scale factor and $N(t)$ is the lapse function. For all models under consideration, we will be able to set $N=1$, which simplifies the cosmological equations further. In this case, $t$ is cosmological time. In this cosmological setting, \eqref{eq:EFE} yields the cosmological Einstein field equations given by
\begin{align}
    3H^2 &= \kappa \Bigl(\rho+\frac12 \dot{\phi}^2+V-6 f H \dot{\phi}\Bigr)\,, \label{eq:fried1}\\
    3H^2+2\dot{H} &= -\kappa\Bigl( p+\frac{1}{2} \dot{\phi}^2- V+2f \ddot{\phi}+2\dot{\phi}^2 \frac{\partial f}{\partial \phi} \Bigr)\,, \label{eq:fried2}
\end{align}
and \eqref{eq:modifiedKG} leads to the modified Klein-Gordon (KG) equation
\begin{equation}
    \ddot{\phi} + 3H\dot{\phi} + \frac{\partial V}{\partial\phi} - 6f(3H^2+\dot{H}) + 18nH^2 \frac{\partial f}{\partial n}=0 \,.
    \label{eq:KG}
\end{equation}
Here the dot denotes differentiation with respect to cosmological time, and we remark that $\rho$ and $p$ are the energy-momentum density and pressure of the fluid, respectively.

A direct, but lengthy, calculation verifies that the three equations~(\ref{eq:fried1})--(\ref{eq:KG}) imply the fluid's energy-momentum conservation equation  $\dot{\rho}+3H(\rho+p)=0$. This is a non-trivial result which is, perhaps, unexpected given that the coupling contains an unspecified function. Let us also note that the only dependence on the scale factor $a(t)$ in the field equations is via the Hubble function $H$ and its derivative. These equations feature both first and second derivatives of the scalar field, $\phi$. However, following~\cite{Copeland:1997et}, one can introduce a new variable which depends on the first derivative of the scalar field, leading to field equations which are first order. In short, this is the key idea behind the dynamical systems formulation.

If we assume that the potential is non-negative, we can introduce the well-known dimensionless variables, first proposed in \cite{Copeland:1997et},
\begin{equation}
    x=\frac{\sqrt{\kappa } \dot{\phi}}{\sqrt{6} H}\,,\quad 
    y=\frac{\sqrt{\kappa V}}{\sqrt{3} H} \,,\quad 
    \sigma=\frac{\sqrt{\kappa \rho}}{\sqrt{3} H}\,.
    \label{eq:variables}
\end{equation}
We restrict to the case that $H>0$, i.e.\ that the universe is expanding (choosing $H<0$ would correspond to a contracting universe). It follows that the variables $y$ and $\sigma$ are non-negative. 

In line with previous studies, we assume $V$ has the exponential form
\begin{equation}\label{eq:exppot}
    V(\phi)=V_0\exp\left(-\kappa\lambda\phi\right) \,,
\end{equation}
where $V_0>0$ is a constant and $\lambda \geq 0$ is a dimensionless parameter. We note that this form for $V$ is invertible, which will allow us to view $\phi$ as a function of $V$. This potential is most convenient as the exponential form allows one to close the autonomous system of equations without the introduction of an additional variable. 

When the FLRW metric is considered in~(\ref{varwrtvarphi}) and~(\ref{varwrttheta}), one immediately finds that the entropy density $s=s_0$ is a constant. Consequently, the coupling function is of the form $f(n,\phi)$ only. Moreover, (\ref{varwrtvarphi}) and~(\ref{varwrttheta}) also imply that the particle number density is $n=n_0 a^{-3}$, where $n_0$ is a constant, which is expected.

Going back to Brown's formulation~(\ref{lagrangianmatter}), we have that the energy density is a function of $n$, since the fluid's entropy $s$ is constant, thus $\rho=\rho(n)$. On the other hand, in standard cosmology, it is customary to assume a linear equation of state of the form $p=w\rho$. We will now demonstrate that, given the definition of pressure in \eqref{eq:defofpressure}, this is equivalent to the assumption that the density is a power of the particle number density. To begin with, let us consider $\rho = n^{w+1}$, for some $w$, which implies
\begin{equation}
    p=n\frac{\partial \rho}{\partial n}-\rho = n (w+1) n^w - n^{w+1} =
    (w+1) n^{w+1} - n^{w+1} = w n^{w+1} = w\rho.
\end{equation}
For the matter dominated case, $w=0$, this gives that $p=0$. On the other hand, integrating \eqref{eq:defofpressure} with the assumption that $p=w\rho$ implies $\rho=n^{w+1}$.

Dividing equation \eqref{eq:fried1} by $3H^2$ and using the variables~(\ref{eq:variables}), one obtains
\begin{equation}
    x^2 + y^2 + \sigma^2 -2\sqrt{6\kappa}\, x f(n,\phi) = 1 \,. 
    \label{eq:friedconstr}
\end{equation}
We note that $f$ must be chosen to have the same dimensions as $\kappa^{-1/2}$ to ensure that this equation is consistent. As it is derived from the Friedmann constraint equation, we will generally refer to it as the \emph{constraint equation}. This is motivated by the fact that it is an algebraic relation between all the variables, which implies that the variables are not all independent.

We finish this Section by noting that for $f(n,\phi)=0$, we retrieve the model studied in~\cite{Copeland:1997et}, which we can view as our baseline model. When interpreting our results, we draw analogies and highlight differences with this baseline model. In that work, the constraint equation~(\ref{eq:friedconstr}) is solved for the matter variable $\sigma$ which is then eliminated from the other equations, reducing the system to two differential equations and we follow the same approach here. 

\section{Constant interaction}
\label{sec:constantint}

To begin our study, we consider perhaps the simplest non-trivial model, where the coupling function is a constant 
\begin{equation}
    f(n,s,\phi)= \frac{k}{\sqrt{24\kappa}}\,,
\end{equation}
for some constant $k$. This model shares some similarities with~\cite{Copeland:1997et} and is an ideal prelude to the study of more complicated models. 

\subsection{General properties and dynamical systems formulation}

Let us start with the Klein-Gordon equation~(\ref{eq:KG}), which simplifies to 
\begin{equation}
    \ddot{\phi}+3H\dot{\phi}+\frac{\partial V}{\partial\phi}=\sqrt{\frac{3}{2}} \frac{k}{\sqrt{\kappa}} \left(3H^2+\dot{H}\right)=:Q \,,
    \label{eq:KG2}
\end{equation}
where we introduce the quantity $Q$ to match previous work on dark sector couplings~\cite{Boehmer:2008av}. The energy density and pressure of the scalar field are given by $\rho_\phi=\dot{\phi}^2/2+V$ and $p_\phi=\dot{\phi}^2/2-V$, respectively. This allows us to re-write equation \eqref{eq:KG2} in the well-known form
\begin{equation}
    \dot{\rho}_\phi+3H(\rho_\phi+p_\phi)=
    \sqrt{\frac{3}{2}} \frac{k}{\sqrt{\kappa}}\dot{\phi}(3H^2+\dot{H})=
    Q\dot{\phi}\,,
    \label{eq:KG3}
\end{equation}
hence $Q$ can be re-expressed as
\begin{equation}
    Q=H^2 \sqrt{\frac{3}{2}} \frac{k}{\sqrt{\kappa}}\left(3+\frac{\dot{H}}{H^2}\right)=
    H^2 \sqrt{\frac{3}{2}} \frac{k}{\sqrt{\kappa}} (2-q),
    \label{eq:KG4}
\end{equation}
where $q=-1-\dot{H}/H^2$ is the standard deceleration parameter. It is well known from dark sector coupling models~\cite{Boehmer:2008av,Tamanini:2015iia} that $Q>0$ means an energy transfer from dark matter to dark energy and $Q<0$ a transfer in the opposite direction.


For $k>0$, equation~(\ref{eq:KG4}) implies that an epoch of accelerated expansion, $q<2$, gives a positive coupling, leading to energy going into the scalar field. In turn, this leads to an epoch of further acceleration and can be seen as a self-reinforcing effect. The above argument is reversed for $f<0$ (i.e.\ $k<0$). Given that the late-time universe is dark energy dominated while the early universe contains considerably more dark matter than dark energy, it is reasonable to consider $f > 0$ (i.e.\ $k>0$) and it will turn out that such models indeed evolve into epochs of late-time accelerated expansion.

Next, we consider a fluid with equation of state $p=w \rho$, which, as discussed at the end of Section~\ref{sec:cosmology}, is equivalent to setting $\rho = n^{1+w}$. The constraint equation~\eqref{eq:friedconstr} then reads
\begin{equation}
    x^2-kx+y^2+\sigma^2=1\,. 
    \label{eq:friedconstr1}
\end{equation}

The quantity $\sigma^2$ is the relative energy density of matter, sometimes denoted by $\Omega_\mathrm{m}$ when discussing explicit cosmological models. For a scalar field, it is helpful to introduce the equation of state
\begin{equation}
    w_\phi=\frac{p_\phi}{\rho_\phi}=\frac{\frac{1}{2}\dot{\phi}^2-V}{\frac{1}{2}\dot{\phi}^2+V} \,.
\end{equation}
Here, $w_\phi\in[-1,1]$ and we get $w_\phi=-1$ when $\dot{\phi}=0$, as is expected for dark energy. The energy density of the scalar field is given by
\begin{equation}
    \Omega_\phi:=x^2+y^2 \,.
\end{equation}
Hence, \eqref{eq:friedconstr1} can also be written as 
\begin{equation}
    \Omega_{\rm m} +\Omega_\phi - k x= 1 \,.
\end{equation}

At this point, it is clear that one can introduce improved variables by completing the square of the $x$-term in \eqref{eq:friedconstr1}. Namely, we write
\begin{equation}
    \left(x-\frac{k}{2}\right)^2+y^2+\sigma^2=1+\frac{k^2}{4},
    \label{eq:friedconstr1b}
\end{equation}
and now divide by the new right-hand side so that we arrive at
\begin{equation}
    X^2 + Y^2 + \Sigma^2 = 1\,,
    \label{eq:sphere}
\end{equation}
where
\begin{equation}
    X = \frac{x-k/2}{\sqrt{1+k^2/4}}\,, \quad
    Y = \frac{y}{\sqrt{1+k^2/4}}\,, \quad
    \Sigma = \frac{\sigma}{\sqrt{1+k^2/4}}\,.
     \label{eq:varnorm}
\end{equation}
These variables will prove particularly useful for our subsequent qualitative analysis. Using equations \eqref{eq:fried1}--\eqref{eq:KG}, one can obtain the acceleration equation 
\begin{equation}
   1+q = -\frac{\dot{H}}{H^2} = \frac{3}{2}\left[ 
   (1+w) - (w-1) X^2 - Y^2\left((1+w)-\frac{\lambda k}{\sqrt{6}} \right)\right]\,,
   \label{eq:acceleration}
\end{equation}
which can be integrated to find $a(t)$ at any given fixed point $(X_0,Y_0)$. The right-hand side, at a fixed point, is constant. If this constant is non-zero, it is straightforward to show that the scale factor evolves as a power law in cosmological time, that is, $a\propto \left(t-t_0\right)^\mu$, where $\mu$ is that power. We therefore have that $\mu$ is given by
\begin{equation}
    \frac{1}{\mu} = \frac{3}{2}\left[ 
   (1+w) - (w-1) X_0^2 - Y_0^2\left((1+w)-\frac{\lambda k}{\sqrt{6}} \right)\right]=1+q \,. 
    \label{eq:powerlaw}
\end{equation}
Here $t_0$ is an integration constant. When the right-hand side of~(\ref{eq:acceleration}) vanishes at some fixed point, the scale factor $a(t)$ evolves exponentially. This corresponds to $H$ being constant at this point, that is, a universe undergoing a de Sitter expansion.

It can be useful to define the total energy density and total pressure of the cosmological model 
\begin{align}
    \widetilde{\rho} &= \rho+\frac{1}{2}\dot{\phi}^2+V+\rho_{\mathrm{int}}\,,\\
    \widetilde{p} &= p+\frac{1}{2}\dot{\phi}^2-V+p_{\mathrm{int}}\,,
\end{align}
where we set $\rho_{\mathrm{int}}=-kH\sqrt{6/\kappa}+\dot{\phi}$ and $p_{\mathrm{int}}=k\ddot{\phi}/\sqrt{6\kappa}$, as suggested by~(\ref{eq:fried1}) and~(\ref{eq:fried2}). This naturally leads to the effective equation of state parameter $\widetilde{w} =  \widetilde{p}/\widetilde{\rho}$. For power law models, this effective equation of state parameter is directly related to the power $\mu$, and one has
\begin{align}
    \mu = \frac{2}{3(1+\widetilde{w})}\,, 
    \quad \text{or} \quad
    \widetilde{w} = \frac{2}{3\mu}-1 \,.
\end{align}
We note that the power $\mu$, the effective equation of state parameter $\widetilde{w}$, and the deceleration parameter $q$, all encode the same physical information.

Similar to previously studied models, the positivity of the matter variable and equation~(\ref{eq:sphere}) imply that $0 \leq \Sigma \leq 1$, and hence $0 \leq X^2 + Y^2 \leq 1$. Together with the fact that $Y \geq 0$, since we are considering an expanding universe, this means that the phase space for the variables $X$ and $Y$ is a semicircle of radius one.

We are now ready to state the dynamical equations of the system, using the convenient variables defined in~(\ref{eq:varnorm}). This leads to two independent equations
\begin{align}
    X' &= \frac{1}{4} \left[\sqrt{6} \lambda \sqrt{k^2+4}\, Y^2 + X \left(Y^2 \left(\sqrt{6} \lambda  k-6 w-6\right)-6 (w-1)\left(X^2-1\right)\right)\right],
    \label{dyn1} \\[1ex]
    Y' &= -\frac{1}{4} Y \left[\sqrt{6} \lambda  \sqrt{k^2+4}\, X+\left(Y^2-1\right)\left(-\sqrt{6} \lambda  k+6 w+6\right)+6 (w-1) X^2\right].
    \label{dyn2}
\end{align}
Here a prime denotes a derivative with respect to the logarithm of the scale factor $\log(a)$. One can now follow the standard dynamical systems approach to study this system, for a review see~\cite{Bahamonde:2017ize}. We begin with the fixed points of~(\ref{dyn1})--(\ref{dyn2}). We note that these are two polynomial equations of degree three, meaning that one could find up to nine real distinct critical points, by Bézout's theorem. If $Y=0$, the second equation is automatically satisfied, and this leads to the solutions $X\in\{-1, 0, +1\}$. Next, excluding $Y=0$, one notes that $Y$ appears only as $Y^2$ in the equations, meaning that there are up to four more solutions. Two of these are at negative values of $Y$, which we exclude, again because we are considering an expanding universe ($Y\ge0$). Assuming $\lambda > 0$ and $-1 \leq w \leq 1$, we obtain a total of five critical points, shown in Table~\ref{tab:constcritp}. 

\begin{table}[!htb]
\makegapedcells
\centering
    \begin{tabular}{|c|c|c|l|} \hline
        Point &$X$ & $Y$ & Existence \\ \hline\hline
        $O$ & $0$ & $0$ & for all $k$  \\ \hline
        $A_{-}$& $-1$ & $0$ & for all $k$ \\ \hline
        $A_{+}$& $1$ & $0$ & for all $k$   \\ \hline
        $B$ & $ \displaystyle 
        \frac{\lambda \sqrt{k^2+4}}{2 \sqrt{6}-\lambda k}$ & 
        $ \displaystyle \frac{2 \sqrt{6-\sqrt{6}k\lambda-\lambda^2}}{2 \sqrt{6}-\lambda k}$ & 
        for $\displaystyle  k < \frac{6-\lambda^2}{\sqrt{6}\lambda}$; note $X^2+Y^2=1$
        \\[1ex] \hline
        $C$ & 
        $\displaystyle \frac{\sqrt{6}(1+w)-\lambda  k}{\lambda \sqrt{k^2+4}}$ & 
        $\displaystyle \frac{\sqrt{\left(\sqrt{6} \lambda k - 6(1+w)\right)(w-1)}}{\lambda \sqrt{k^2+4}}$ & 
        \makecell[l]{
        for $-1 \leq w < 1$, \\[1ex] 
        $\displaystyle 2\sqrt{\frac{2}{3}}\frac{3(1+w)-\lambda^2}{(3+w)\lambda} \leq k \leq \sqrt{6}\frac{1+w}{\lambda}$ \\[3ex]
        for $w=1$, \quad $\displaystyle k \geq \frac{6-\lambda^2}{\sqrt{6}\lambda}$
        }
        \\[1ex] \hline
    \end{tabular}
   \caption{Critical points of system~(\ref{dyn1})--(\ref{dyn2}).}
   \label{tab:constcritp}
\end{table}
Note that Point $B$ is always located on the boundary of the phase space while Point $C$ is generally inside the phase space, if it exists. For the special value \begin{equation}
    k = 2\sqrt{\frac{2}{3}}\frac{3(1+w)-\lambda^2}{(3+3)\lambda}\,,
\end{equation}
the lower existence bound, Point $C$ is also on the boundary. Next, one needs to study the eigenvalues of the stability matrix at each of the critical points. For more details about their classification, see Appendix~\ref{sec:AppendixA}. For the first four points, $O$, $A_{\pm}$, and $B$, these are given in Table~\ref{tab2}. Note that we will discuss the occurrence of possible zero eigenvalues separately to keep the discussion more straightforward. For example, one may immediately note that the choice $w=1$ implies at least one zero eigenvalue for the Points $O$ and $A_{\pm}$.

\begin{table}[!htb]
\makegapedcells
\centering
    \begin{tabular}{|c|l|l|} \hline 
         Point & Eigenvalues & Classification  \\\hline\hline 
         $O$ & 
         $\displaystyle \frac{3}{2}(w-1)$, \quad 
         $\displaystyle \frac{3}{2}(1+w)-\frac{\sqrt{6}}{4}k\lambda $ & 
         \makecell[l]{
         saddle if $k\lambda < \sqrt{6}(1+w)$ \\[1ex]
         stable if $k\lambda > \sqrt{6}(1+w)$
         } \\\hline 
         $A_{-}$ & 
         $3(1-w)$,\quad 
         $\displaystyle 3-\sqrt{\frac{3}{8}}\lambda\left(k-\sqrt{k^4+4}\right)$ & 
         saddle for all $\lambda$
         \\\hline 
         $A_{+}$ & $3(1-w)$,\quad 
         $\displaystyle 3-\sqrt{\frac{3}{8}}\lambda\left(k+\sqrt{k^4+4}\right)$ & 
         \makecell[l]{
         saddle if $\displaystyle k < \frac{6-\lambda^2}{\sqrt{6}\lambda}$\\[3ex] 
         stable if $\displaystyle k > \frac{6-\lambda^2}{\sqrt{6}\lambda}$
         }\\\hline 
         $B$ & 
         $\displaystyle \frac{2 \sqrt{6} \left(\lambda ^2+6\right)}{2 \sqrt{6}-\lambda  k}-3 (w+3)$,\quad $\displaystyle \frac{\sqrt{6}(6+\lambda^2)}{2 \sqrt{6}-\lambda  k}-6$ & 
         \makecell[l]{
         unstable if $\displaystyle \frac{6-\lambda^2}{\sqrt{6}\lambda} < k < \frac{2\sqrt{6}}{\lambda}$\\[3ex] 
         saddle if $\displaystyle 2\sqrt{\frac{2}{3}}\frac{3(1+w)-\lambda^2}{(3+w)\lambda} < k < \frac{6-\lambda^2}{\sqrt{6}\lambda}$ \\[3ex]
         stable if $\displaystyle k > \frac{2\sqrt{6}}{\lambda}$ or 
         $\displaystyle k < 2\sqrt{\frac{2}{3}}\frac{3(1+w)-\lambda^2}{(3+w)\lambda}$
         }
        \\\hline 
    \end{tabular}
\caption{Stability of the critical Points $O$, $A_{\pm}$ and $B$ for system~(\ref{dyn1})--(\ref{dyn2}). The classification assumes that the eigenvalues are non-zero.}
\label{tab2}
\end{table}

The final critical point, $C$, is more difficult to study as the eigenvalues are much more involved. They are the solutions of the characteristic polynomial in $\xi$
\begin{multline}
    0 = \xi^2 + \frac{3}{2}(1-w)\xi -\frac{3(1-w)}{2(k^2+4)\lambda^2}
    \\ \times
    \Bigl(3 \lambda ^2 k^2 (w+3)+\sqrt{6} \lambda  k (2 \lambda ^2-3 (w+1) (w+5))+12 (w+1) (-\lambda ^2+3 w+3)\Bigr) \,.
\end{multline}
Solving this quadratic equation is easy, however, the explicit solutions do not offer much insight given that they contain three free parameters. For concrete parameter choices, we discuss this point in more detail below. One easy result to extract is the sum of the eigenvalues $\xi_1$ and $\xi_2$ at this point, that is,
\begin{align}
    \xi_1 + \xi_2 = -\frac{3}{2}(1-w) \,.
\end{align}
As this number is negative for $w < 1$, this point cannot have two positive eigenvalues and therefore will have at least one stable direction. This implies that Point $C$ is a saddle point, stable node, or stable spiral.

There will be many parameter choices resulting in zero eigenvalues, $w=1$ being the obvious one. However, the choice $k\lambda=\sqrt{6}(1+w)$ would also give a zero eigenvalue for Point $O$. The stability analysis of such points requires techniques beyond linear stability theory. These are well known and their applications in cosmology were discussed in~\cite{Boehmer:2011tp,B_hmer_2016,Bahamonde:2017ize}. However, for the purposes of this work, we will assume a matter dominated universe $w=0$ and employ linear stability theory. We note that our analysis can also be performed for the radiation dominated case $w=1/3$ where one finds qualitatively similar results.

\subsection{The matter dominated case}

In what follows, we set $w=0$. Points $O$, $A_{\pm}$ and $B$ are independent of $w$ and all results discussed above apply. The location of Point $C$, if it exists, depends on $w$ and so do its corresponding eigenvalues. We now outline some physical properties of the critical points of the system, with the values of effective equation of state parameter $\widetilde{w}$ and the deceleration parameter shown in Table~\ref{tab3}.

\begin{table}[!htb]
\makegapedcells
    \centering
    \begin{tabular}{|c|c|c|} \hline
    Point & $\widetilde{w}$ & $q$  \\ \hline\hline  $O$ & $0$ & $1/2$\\\hline
    $A_{-}$& $1$ & $2$\\ \hline
    $A_{+}$& $1$ & $2$\\ \hline
    $B$ & $\displaystyle \frac{2 \sqrt{6} \left(\lambda ^2-3\right)+9 k \lambda }{6 \sqrt{6}-3 k \lambda}$ & $\displaystyle \frac{\sqrt{6} \left(\lambda ^2-2\right)+4 k \lambda }{2 \sqrt{6}-k \lambda }$\\ \hline
    $C$ & $0$ & $1/2$\\ \hline
    \end{tabular}
    \caption{Physical properties of the fixed points for the matter dominated case, for system~(\ref{dyn1})--(\ref{dyn2}).}
    \label{tab3}
\end{table}

\begin{figure}[!htb]
    \centering
    \includegraphics[width=0.80\textwidth]{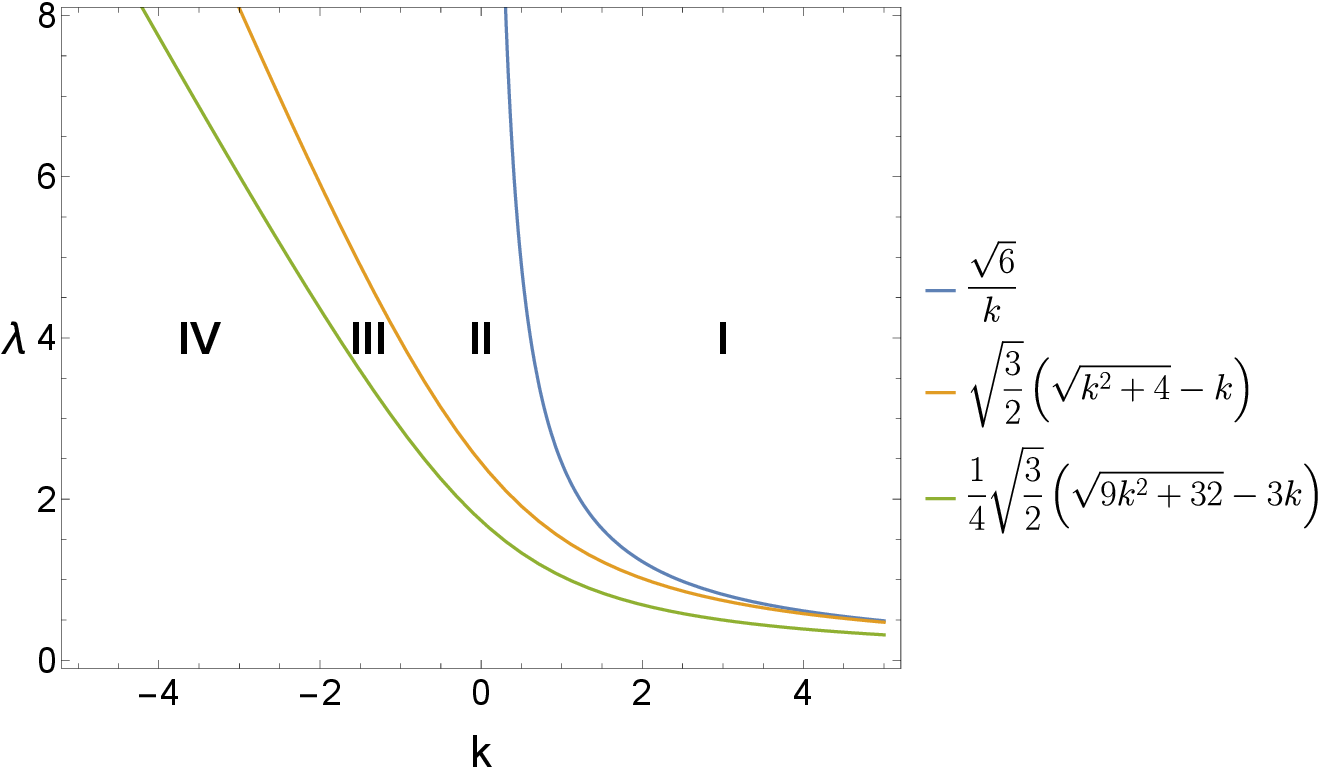}
    \caption{Existence and stability regions in $(k,\lambda)$-plane. The plotted curves follow from the stability criteria shown in Table~\ref{tab2}.}
    \label{fig:regions}
\end{figure}

In order to analyse the stability of the fixed points, we look at the different regions in the $(k,\lambda)$-plane, see Fig.~\ref{fig:regions}, and we recall that $\lambda >0$. First, we remark that the fixed Points $O$ and  $A_{\pm}$ exist for all values of $\lambda$ and $k$. Moreover, there are four distinct regions of values of $k$ and $\lambda$, which yield different stability properties of the critical points, and, hence, different cosmological phenomena. We discuss these four different cases and comment on their suitability as a cosmological model. 

\paragraph{Region I.} For values within Region I, there are only three critical points, namely $O$ and $A_{\pm}$. In particular, $O$ is a stable node, $A_{-}$ is an unstable node, and $A_{+}$ is a saddle. Since $O$ is the only attractor of the system, all trajectories will eventually approach it. The Point $A_{-}$ can be thought of as the past-time attractor, in the sense that all trajectories would start at $A_{-}$. Lastly, some trajectories are attracted towards $A_{+}$, but are eventually repelled and move towards $O$. This case is not of physical interest. We do not show a phase space diagram.

\begin{figure}[!htb]
    \centering
    \includegraphics[width=0.95\textwidth]{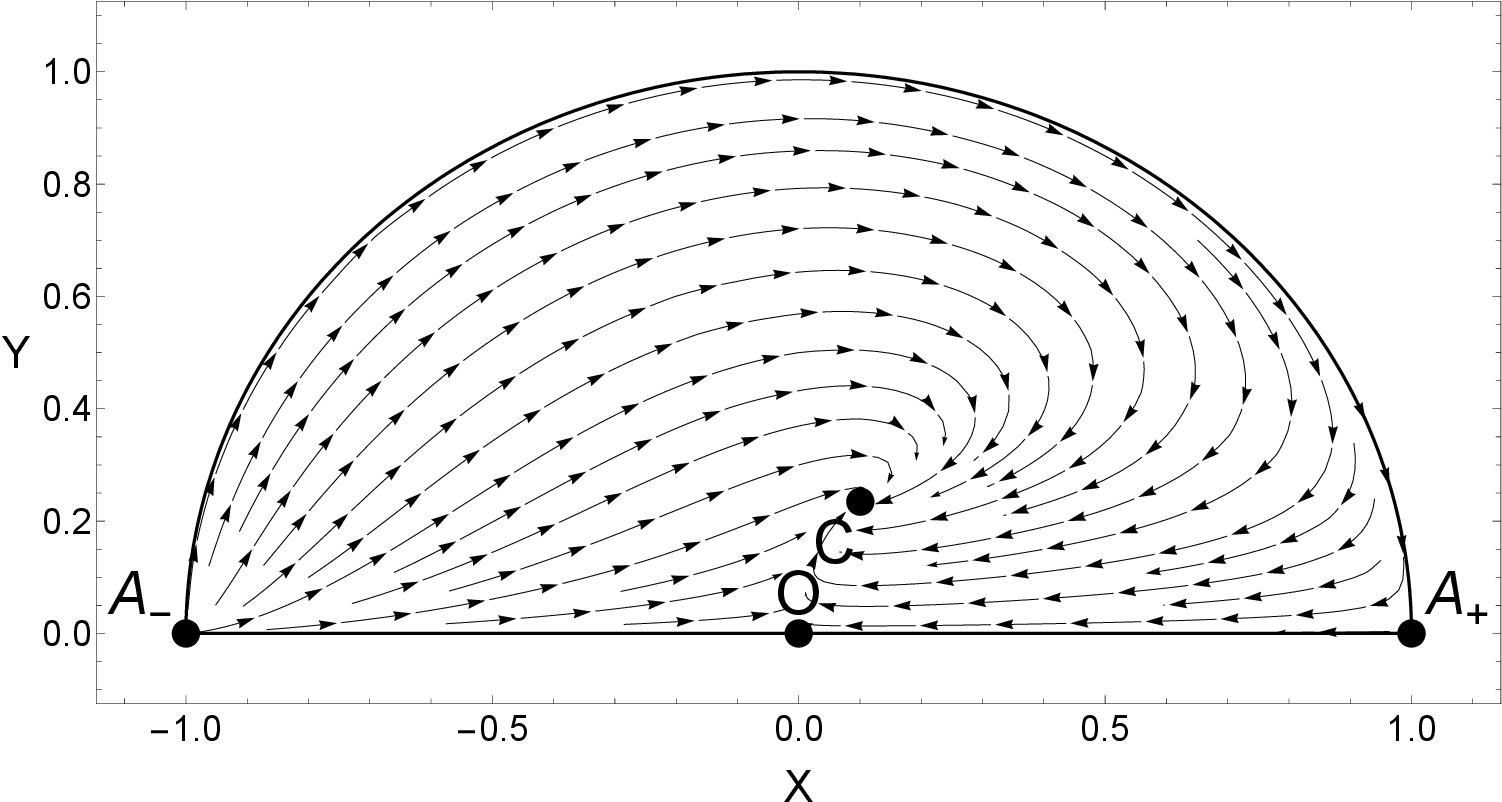}
    \caption{Phase space with $k=1$ and $\lambda=2$. $B$ is a stable node, that is, the only attractor describing a scaling solution with $\widetilde{w}=w=0$. No acceleration region present.}
    \label{fig:constfunctregion2}
\end{figure}

\paragraph{Region II.} In Region II, Point $B$ does not exist. We therefore have four critical points: the unstable node at Point $A_{-}$, the stable node at Point $C$, and the saddle Points $O$ and $A_{+}$. We note that here, Point $C$ represents the \emph{scaling solution}~\cite{Amendola:1999qq} as the effective equation of state parameter matches the matter one ($\widetilde{w}=w=0$). Hence, the universe expands as if it was completely matter dominated despite the scalar field's influence, according to~\eqref{eq:powerlaw}. We note that this is not an accelerated expansion but this solution is of physical relevance for the coincidence problem. The type of dynamics is illustrated by Fig.~\ref{fig:constfunctregion2}, where we set $k=1$ and $\lambda=2$.  We remark the analogy with one of the cases discussed in~\cite{Copeland:1997et}, however we also point out that, in our example, no acceleration region is present.  

\begin{figure}[!htb]
    \centering
    \includegraphics[width=0.95\textwidth]{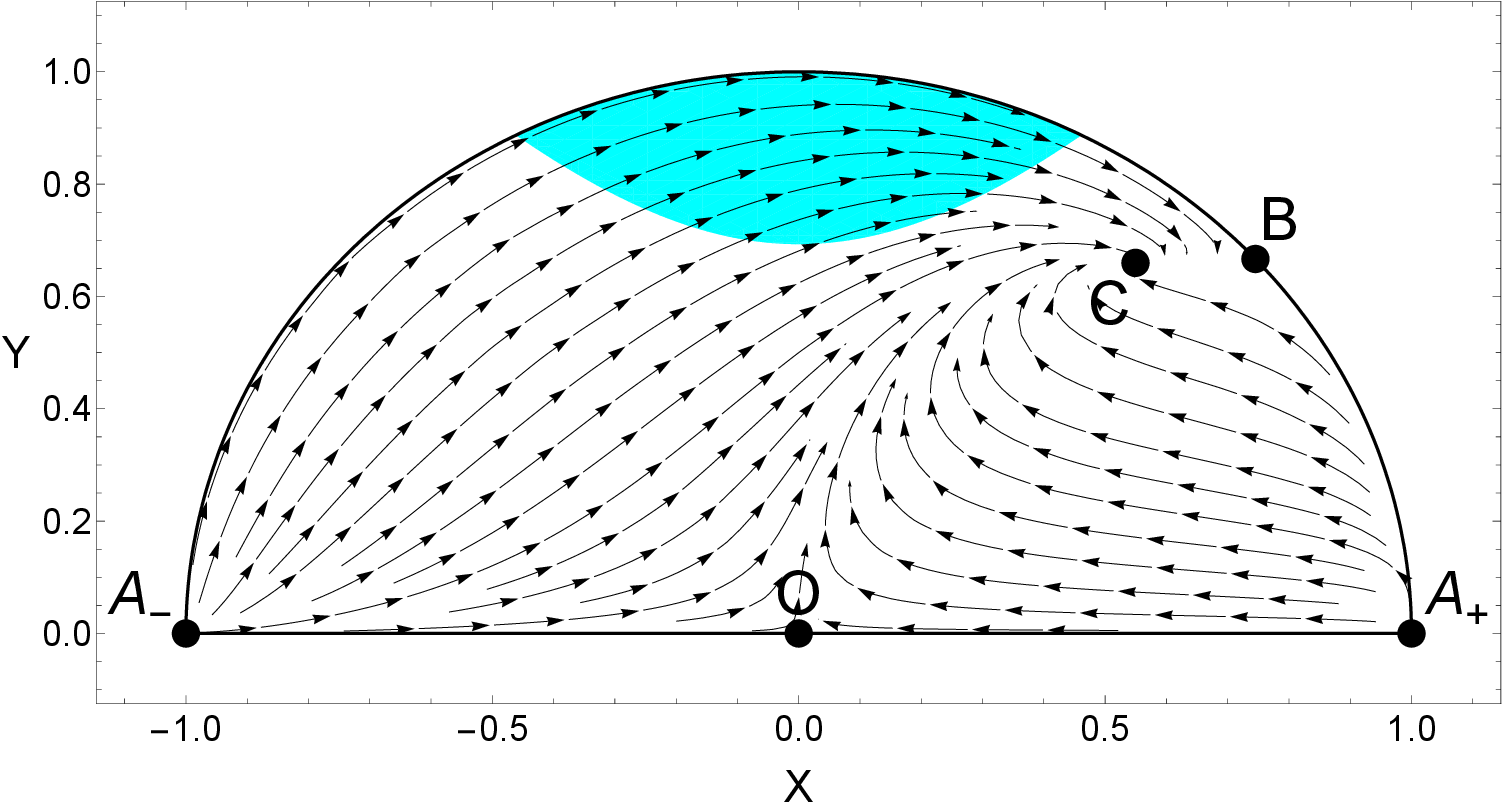}
    \caption{Phase space with $k=1/2$ and $\lambda=3/2$. Here the only attractor is Point $C$ where the universe expands as if it is completely matter dominated (scaling solution), while Point $B$ is a saddle point. The shaded region represents the area of the phase space where there is accelerated expansion.}
    \label{fig:constfunctregion3}
\end{figure}

\paragraph{Region III.} In Region III, there are five critical points in the phase space. Points $A_{-}$, $O$, and  $C$, still behave as an unstable node, a saddle point, and a stable node, respectively. Point $A_{+}$ is now an unstable node. Point $B$ exists and is a saddle point. This is shown in Fig.~\ref{fig:constfunctregion3}, where $k=1/2$ and $\lambda=3/2$. Point $C$ always lies outside the acceleration region, so it does not represent a late-time inflationary solution. This is, again, a scaling solution. We highlight the analogy with another case discussed in~\cite{Copeland:1997et}. All trajectories connect Points $A_{\pm}$ to Point $C$, with the exception of the orbits along the boundary. 

\begin{figure}[!htb]
    \centering
    \includegraphics[width=0.95\textwidth]{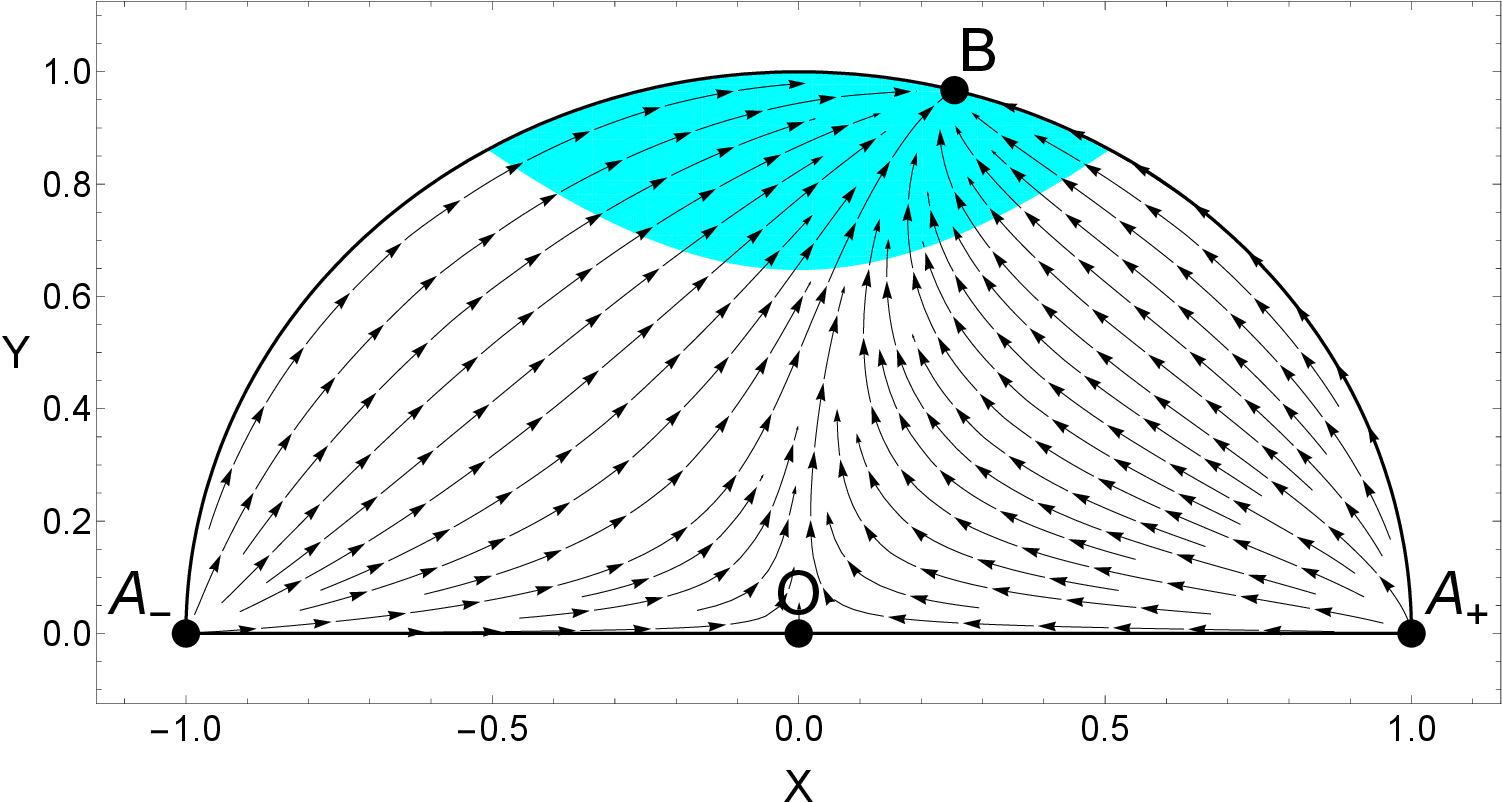}
    \caption{Phase space with $k=1$ and $\lambda=1/2$. Here Point $B$ is the only attractor and represents a late-time inflationary cosmological solution. The shaded area represents the part of the phase space where there is accelerated expansion.}
    \label{fig:constfunctregion4}
\end{figure}

\paragraph{Region IV.} In Region IV, there are again four fixed points since Point $C$ lies outside the physical space. Here, Point $A_{-}$ is always an unstable node and can be seen as the past attractor. Similar to Region III, $A_{+}$ is an unstable node. Point $O$ is a saddle point, whereas Point $B$ is a stable node and therefore the late-time attractor. We note that here Point $B$ lies within the region of accelerated expansion, hence we are in the presence of a cosmological solution with accelerated expansion. This is illustrated in Fig.~\ref{fig:constfunctregion4}. Once again, we emphasise the analogy with one of the cases discussed in~\cite{Copeland:1997et}.

\section{A non-constant interaction model}
\label{sec:nonconstint}

\subsection{Equations of the model}

We are now considering a model with a non-constant interaction term. As we wish to exploit dynamical systems techniques without increasing the number of independent variables, we consider an interaction of the form~\cite{Nunes:2000ka,Teixeira:2019tfi}
\begin{align}
    f(n,\phi)=\frac{k}{2\sqrt{6\kappa}}n^{\alpha/(2(1+w))}
    V^{-\alpha/2} \,,
\end{align}
where $\alpha$ is a fixed power. Let us make the following observations to motivate this particular choice for $f$. 
In cosmology, $s=s_0$ and $n=n_0 a^{-3}$. Moreover, if we consider the linear equation of state $p=w\rho$ and the definition of $p$ from Brown's fluid model, one has $\rho = n^{w+1}$. Therefore, this specific form allows us to use the variables introduced in~\eqref{eq:variables} directly, so~\eqref{eq:friedconstr} becomes
\begin{align}
    1-x^2-y^2-\sigma^2+k\,x \frac{\sigma^\alpha}{y^{\alpha}}=0 \,.
    \label{fried-non1}
\end{align}
The dynamics of the system depends on the parameters $w$, $\lambda$, $k$, and $\alpha$. We note that one could consider the limit $\alpha \to 0$ and recover equation~(\ref{eq:friedconstr1}). 

It is clear that larger (integer) values of $\alpha$ can make the study of this system difficult, since the constraint~(\ref{fried-non1}) would become a polynomial equation of high order. At the same time, even the values $\alpha = \pm 1$ introduce challenges as one has to deal with cubic equations. In fact, the two simplest cases that can be studied explicitly, without introducing further complications, are $\alpha=\pm 2$. In what follows, we consider $\alpha=2$ and $w=0$, and note that the radiation dominated case ($w=1/3$) leads to broadly similar results.

When $\alpha=2$, the constraint~(\ref{fried-non1}) can be written as 
\begin{align}
    &y^2\left(1-x^2-y^2-\sigma^2\right)+k\,x \sigma^2 = 0 \,,
   \nonumber \\ \Leftrightarrow \quad
   &y^2\left(1-x^2-y^2\right) + \sigma^2 \left(kx-y^2\right) = 0 \,,
   \nonumber \\ \Leftrightarrow \quad
   &\sigma^2 = \frac{y^2\left(1-x^2-y^2\right)}{y^2-kx} \,,
    \label{fried-non2}
\end{align}
allowing us to eliminate $\sigma$ from the equations, so that the dynamical system remains two-dimensional. Moreover, as $\sigma^2 \geq 0$, we have
\begin{align}
    \frac{y^2(1-x^2-y^2)}{y^2-kx} \geq 0 \,.
    \label{fried-non2a}
\end{align}
This gives rise to the physical regions of the phase space
\begin{alignat}{2}
    &1-x^2-y^2 \geq 0 \quad\text{and}\quad &y^2-kx > 0 \,,\label{fried-non2bi}\\
    &1-x^2-y^2 \leq 0 \quad\text{and}\quad &y^2-kx < 0 \,.
    \label{fried-non2bii}
\end{alignat}

Since the non-constant coupling leads to a considerably more complicated dynamical system, we restrict our study to the matter dominated case $w=0$. The dynamical equation for $x$ is given by
\begin{align}
    x'&=\frac{2 y^4 \left(3 x^3-3 x \left(y^2+1\right)+\sqrt{6} \lambda  y^2\right)-k\,\mathcal{A}+k^2\,\mathcal{B}}{4 y^4-8 k x y^2+k^2 \left(x^4+2 x^2 \left(y^2+1\right)+\left(y^2-1\right)^2\right)}\,,
    \label{noncon1}
\end{align}
where the functions $\mathcal{A}$ and $\mathcal{B}$ are defined by
\begin{align}    
    \mathcal{A} &= y^2 (x^2-1) (x^2 (2 \sqrt{6} \lambda  x+3)-3) +
    3 y^4 (\sqrt{6} \lambda  (x^3+x)-6 x^2-2) +
    y^6 (\sqrt{6} \lambda x+3)\,,
    \label{noncon2}\\
    \mathcal{B}&= x \bigl[\sqrt{6} \lambda  x (x^4+(x^2+3) y^2-1)-
    3 (x^4+4 x^2 y^2-y^4)-6 y^2+3\bigr]\,.
    \label{noncon3}
\end{align}
Similarly, the $y$ equation reads
\begin{align}    
    y' &= \frac{-4 y^5 (-3 x^2+\sqrt{6} \lambda  x+3 y^2-3)-k\,\mathcal{C}+k^2\,\mathcal{D}}{2\left[4 y^4-8 k x y^2+k^2 \left(x^4+2 x^2 \left(y^2+1\right)+\left(y^2-1\right)^2\right)\right]}\,,
    \label{noncon4}
\end{align}
where
\begin{align}
   \mathcal{C}&= 2 y^3 \left(2 \sqrt{6} \lambda  x^4+6 x^3+3 \sqrt{6} \lambda  x^2 \left(y^2-2\right)-18 x
   \left(y^2-1\right)+\sqrt{6} \lambda  y^2 \left(y^2-1\right)\right)\,,
   \label{noncon5}\\
   \mathcal{D}&= x y \left(\sqrt{6} \lambda  \left(3 x^4+x^2 \left(4 y^2-6\right)+y^4-1\right)-24 x
   \left(y^2-1\right)\right)\,.
   \label{noncon6}
\end{align}
Here a prime denotes a derivative with respect to the logarithm of the scale factor $\log(a)$. This way of writing the dynamical equations, namely isolating the terms in powers of $k$, is useful as it allows us to consider the limit $k \to 0$ easily. In that case these equations reduce to those of~\cite{Copeland:1997et}.

We remark that the acceleration equation, which follows from~\eqref{eq:fried1} and \eqref{eq:KG}, is
\begin{equation}
    \frac{\dot{H}}{H^2}=\frac{6 y^4 \left(-x^2+y^2-1\right)+k\,\mathcal{E}+k^2\,\mathcal{F}}{4 y^4-8 k x y^2+k^2\left(x^4+2 x^2 \left(y^2+1\right)+\left(y^2-1\right)^2\right)}
    \label{noncon7}
\end{equation}
where the functions $\mathcal{E}$ and $\mathcal{F}$ are given by
\begin{align}
    \mathcal{E} & =\frac{\mathcal{C}}{2y}+4\sqrt{6}\lambda x^2y^2\,,
   \label{noncon8}\\
   \mathcal{F} &= x \left(12 x \left(y^2-1\right)-\sqrt{6} \lambda \left(x^2+y^2-1\right) \left(2
   x^2+y^2\right)\right) \,.
   \label{noncon9}
\end{align}

\subsection{Critical points and stability}

To find the critical points, we need to solve the equations $x'=0$ and $y'=0$ simultaneously, which is a non-trivial task since both numerators are polynomials of degrees seven, giving up to 49 roots. Many of those will lie outside the physical phase space, while others will come in complex conjugate pairs which also have no physical significance. At this point, it is not clear how many physical critical points this system will have for arbitrary $\lambda$ and $k$ and hence one has to investigate the system carefully to extract them. 

One way to find the critical points is to draw inspiration from the previous model. For example, setting $y=0$ in~\eqref{noncon4} leads to $y'=0$, while setting $y=0$ in~\eqref{noncon1} means that $x'=0$ simplifies to
\begin{align}
    \frac{x\left(x^2-1\right)\left(\sqrt{6}\lambda x -3\right)}{1+x^2} = 0 \,.
    \label{noncon10}
\end{align}
This yields the first set of critical points $(-1,0)$, $(0,0)$, $(1,0)$, and  $(\sqrt{3/2}\,/\lambda,0)$.

Secondly, we investigate critical points on the unit circle. By substituting $x=\cos\theta$ and $y=\sin\theta$ into~\eqref{noncon1} and~\eqref{noncon4}, at a critical point we obtain
\begin{align}
    \frac{1}{2}\left(\sqrt{6}\lambda-6\cos\theta\right)\sin^2\negmedspace\theta &= 0\,,
    \label{noncon12} \\
    \frac{1}{2}\left(\sqrt{6}\lambda-6\cos\theta\right)\sin\theta \cos\theta &= 0\,.
    \label{noncon13}
\end{align}
This gives another critical point at $(\lambda/\sqrt{6},\sqrt{1-\lambda^2/6})$.

\begin{figure}[!htb]
    \centering
    \includegraphics[width=0.6\textwidth]{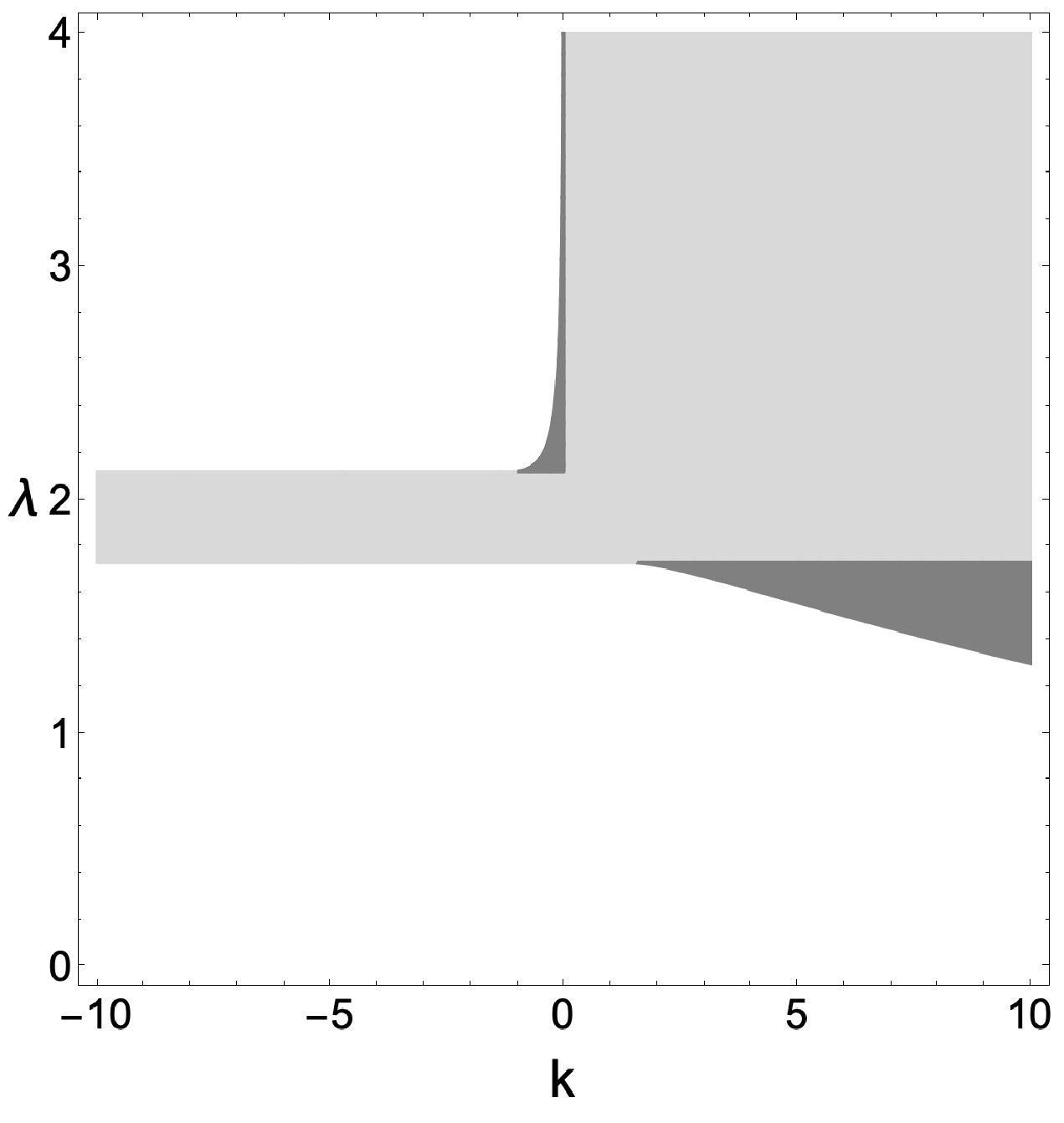}
    \caption{Point $D_{+}$ exists in the whole shaded region. Point $D_{-}$ exists only in the dark grey region.}
    \label{fig:nonex}
\end{figure}

Lastly, one can verify that setting $x_0=\sqrt{3}\,\big/(\sqrt{2}\lambda)$ in the dynamical equations gives four additional solutions, other than $y_0=0$, which are
\begin{align}
    \left(\widehat{y}_{\pm}\right)^2 = \frac{6+\sqrt{6}k\lambda}{8\lambda^2} \pm \frac{1}{8\lambda^2}
    \sqrt{36+2k\lambda\left(3k\lambda+2\sqrt{6}\left(4\lambda^2-15\right)\right)} \,.
    \label{noncon15}
\end{align}
We are not able to find other critical points in the physical phase space, either analytically or numerically. The critical points discussed above are summarised in Table~\ref{noncontable}, together with their corresponding value of the effective equation of state parameter and the value of the deceleration parameter. Note that there will be parameter regions where the critical points with $y$-coordinate $\widehat{y}_{\pm}$, called $D_\pm$, may not exist or where only one of these exits, see Fig.~\ref{fig:nonex}.

\begin{table}[!htb]
\makegapedcells
\centering
    \begin{tabular}{|c|c|c|c|c|c|} \hline
        Point & $x$ & $y$ & $\widetilde{w}$ & $q$ & Existence 
        \\ \hline\hline
        $O$ & $0$ & $0$ & $-1$ & $-1$ & all $\lambda,k$
        \\ \hline
        $A_{-}$& $-1$ & $0$ & $1$ & $2$ & all $\lambda,k$
        \\ \hline
        $A_{+}$& $1$ & $0$ & $1$ & $2$ & all $\lambda,k$  
        \\ \hline
        $B$ & $\lambda/\sqrt{6}$ & $\sqrt{1-\lambda^2/6}$ & $-1+\lambda^2/3$ & $-1+\lambda^2/2$ & $\lambda\leq \sqrt{6}$
        \\ \hline
        $C$ & $\sqrt{3}\,\big/\left(\sqrt{2}\lambda\right)$ & $0$ & $-1 + 12/(3+2\lambda^2)$ & $-1+18/(3+2\lambda^2)$ & $\lambda\neq 0$
        \\ \hline
        $D_{+}$ & $\sqrt{3}\,\big/\left(\sqrt{2}\lambda\right)$ & $\widehat{y}_{+}$ & 0 & $1/2$ & see Fig.~\ref{fig:nonex} 
        \\ \hline
        $D_{-}$ & $\sqrt{3}\,\big/\left(\sqrt{2}\lambda\right)$ & $\widehat{y}_{-}$ & 0 & $1/2$ & see Fig.~\ref{fig:nonex} 
        \\ \hline
    \end{tabular}
    \caption{Critical points of the dynamical system~\eqref{noncon1} and~\eqref{noncon4}, for which an explicit expression could be found.}
    \label{noncontable}
\end{table}

\clearpage

We are now ready to investigate the stability of the critical points. This is straightforward for the Points $O$, $A_{\pm}$, $B$, and $C$. The result are collected in Table~\ref{noncontable2}.

\begin{table}[!htb]
\makegapedcells
\begin{center}
    \begin{tabular}{|c|l|l|} \hline 
         Point & Eigenvalues & Classification \\\hline \hline
         $O$ & $3,\quad 0$ & unstable 
         \\ \hline
         $A_{-}$ & $\displaystyle -\sqrt{6} \lambda -3,\quad \sqrt{\frac32} \lambda +3$ & saddle point ($\lambda > 0$)
         \\ \hline
         $A_{+}$ & $\displaystyle 3-\sqrt{\frac{3}{2}} \lambda ,\quad \sqrt{6} \lambda -3$ & \makecell[l]{unstable node if $\sqrt{3/2}<\lambda <\sqrt{6}$\\[1ex] saddle if $\lambda<\sqrt{3/2}$ or $\lambda>\sqrt{6}$}
         \\ \hline
         $B$ & $\displaystyle \frac{1}{2} \left(\lambda ^2-6\right),\quad \lambda ^2-3$ &\makecell[l]{stable if $\lambda<\sqrt{3}$\\[1ex] saddle if $\sqrt{3}<\lambda <\sqrt{6}$\\[1ex]unstable if $\lambda>\sqrt{6}$}\\\hline
         $C$ & $\displaystyle \frac{18}{2 \lambda ^2+3}-\frac{3}{2},\quad \frac{18}{2 \lambda ^2+3}-3$ & 
         \makecell[l]{
         unstable if $\lambda<\sqrt{3/2}$\\[1ex] 
         saddle if $ \sqrt{3/2}<\lambda <3/\sqrt{2}$\\[1ex]
         stable if $\lambda>3/\sqrt{2}$}\\\hline
    \end{tabular}
\end{center}
\caption{Stability properties of the critical points assuming $\lambda>0$ for the non-constant interaction model.}
\label{noncontable2}
\end{table}

For the Points $D_{\pm}$, the closed form expressions for the eigenvalues are very long and do not offer physical insight. However, when presenting specific cases, we give numerical values for the eigenvalues and discuss the various critical points in more detail. 

Table~\ref{noncontable} suggests that Point $C$ is of particular interest to us. This point has an effective equation of state parameter $\widetilde{w} < -1/3$ if $\lambda < \sqrt{15/2}$, and it is not located on the boundary of the phase space. For such a choice of $\lambda$, we note $\sqrt{15/2} > 3/\sqrt{2}$, which means this point will be an attractor of the dynamical system. In turn, such a model will naturally give rise to a period of late-time accelerated expansion. Moreover, since $\sqrt{15/2} > \sqrt{6}$, Point $B$ will not exists in this case. 

The physical phase space for these models is delineated by the upper semicircle of unit radius centred at the origin and the parabola $x^2=y/k$, as described by \eqref{fried-non2bi} and \eqref{fried-non2bii}. Subsequent figures will make clear which regions form the physical phase space. For all $k$, the semicircle and the parabola will intersect, creating two regions which meet at a point, and which trajectories can traverse. Note that the intersection point is, in general, not a critical point.

\subsection{Phase space diagrams and physical interpretation}

Different choices of $\lambda$ and $k$ result in rather different cosmological models, since the number of critical points and their location vary significantly. Below we consider several cases which illustrate the diversity of the dynamical behaviour exhibited by the model. We select values of $\lambda$ and $k$ systematically, but do not necessarily include every possible scenario which could arise in these models.

\begin{figure}[!htb]
    \centering
    \includegraphics[width=0.95\textwidth]{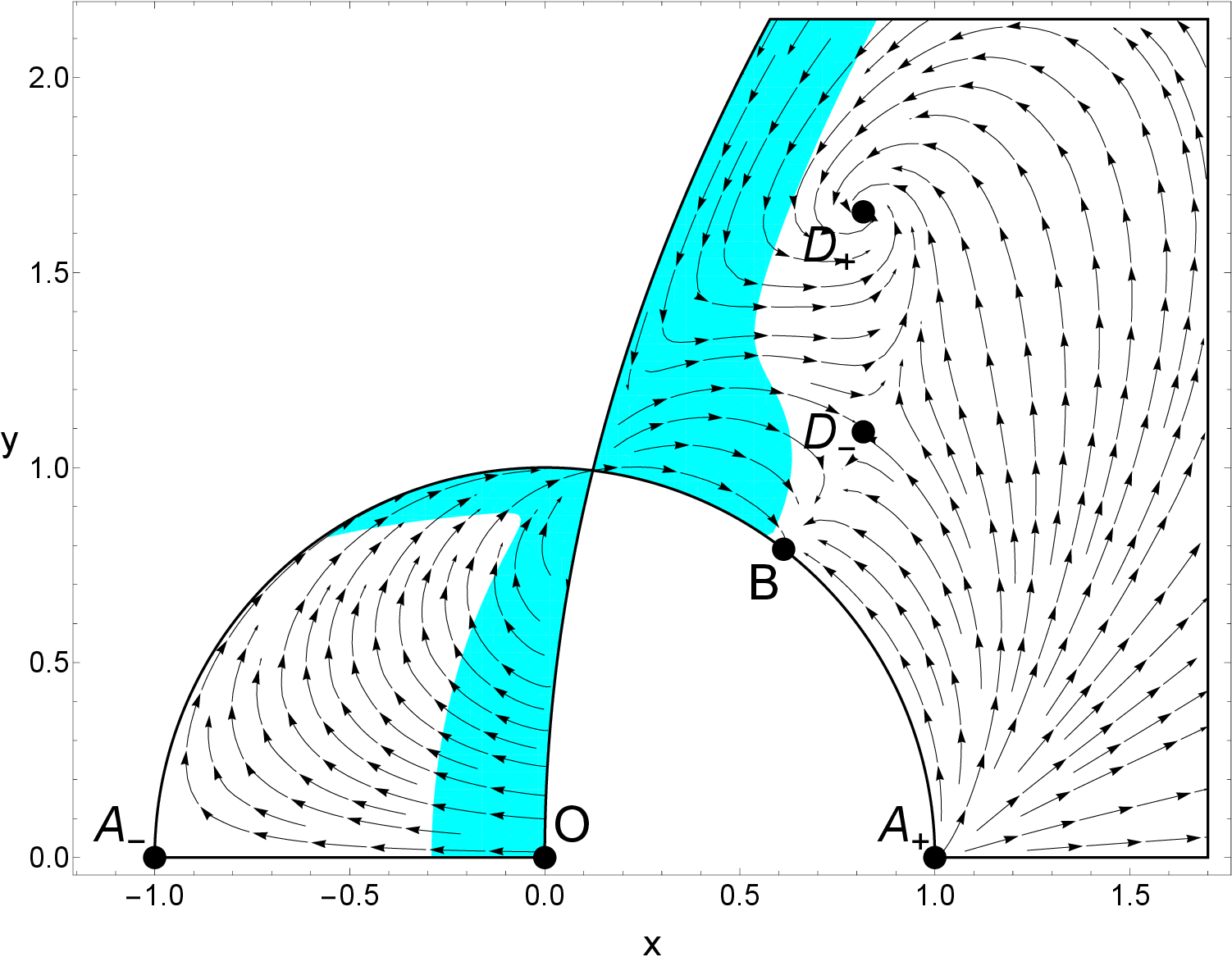}
    \caption{The parameter values are $\lambda=3/2$ and $k=8$. The eigenvalues of $D_{+}$ are $-0.75 \pm 1.3713 \mathrm{i}$ and the eigenvalues of $D_{-}$ are $-2.1570$ and $0.6570$. The shaded area represents the part of the phase space where there is accelerated expansion.}
    \label{fig:many}
\end{figure}

\paragraph{Case (i).} We begin with $\lambda=3/2$ and $k=8$, as shown in Fig.~\ref{fig:many}. In this case, Point $C$ does not exist, however, the other six critical points do exist. The phase space contains a region of accelerated expansion and we note that only Point $O$ is in this region. Point $O$ is an early-time attractor of the phase space, hence, this point could correspond to an early-time universe undergoing accelerated expansion. The other early-time attractor is Point $A_{+}$ which has an effective equation of state parameter $\widetilde{w}=1$, and corresponds to the scalar field's kinetic energy being dominant. Trajectories starting at $O$ will eventually reach the stable Point $B$, where the effective equation of state parameter is $\widetilde{w}=-1/4$. We note that this is negative but not less than $-1/3$, and therefore, not accelerating. Depending on the initial conditions chosen, some trajectories will approach Point $D_{-}$ with $\widetilde{w}=0$, which is matter dominated. On the other hand, trajectories starting at $A_{+}$ will either also terminate at $B$, or reach $D_{+}$. This latter point is a stable spiral with $\widetilde{w}=0$, and hence corresponds to a matter dominated universe. It is interesting to note that trajectories in this case can briefly go through a region of accelerated expansion before reaching $D_{+}$. While these parameter values yield an interesting phase space with a rich structure, this specific model has limited applicability for modern cosmology, since the stable fixed points do not lie within the accelerated region of the phase space.

\begin{figure}[!htb]
    \centering
    \includegraphics[width=0.95\textwidth]{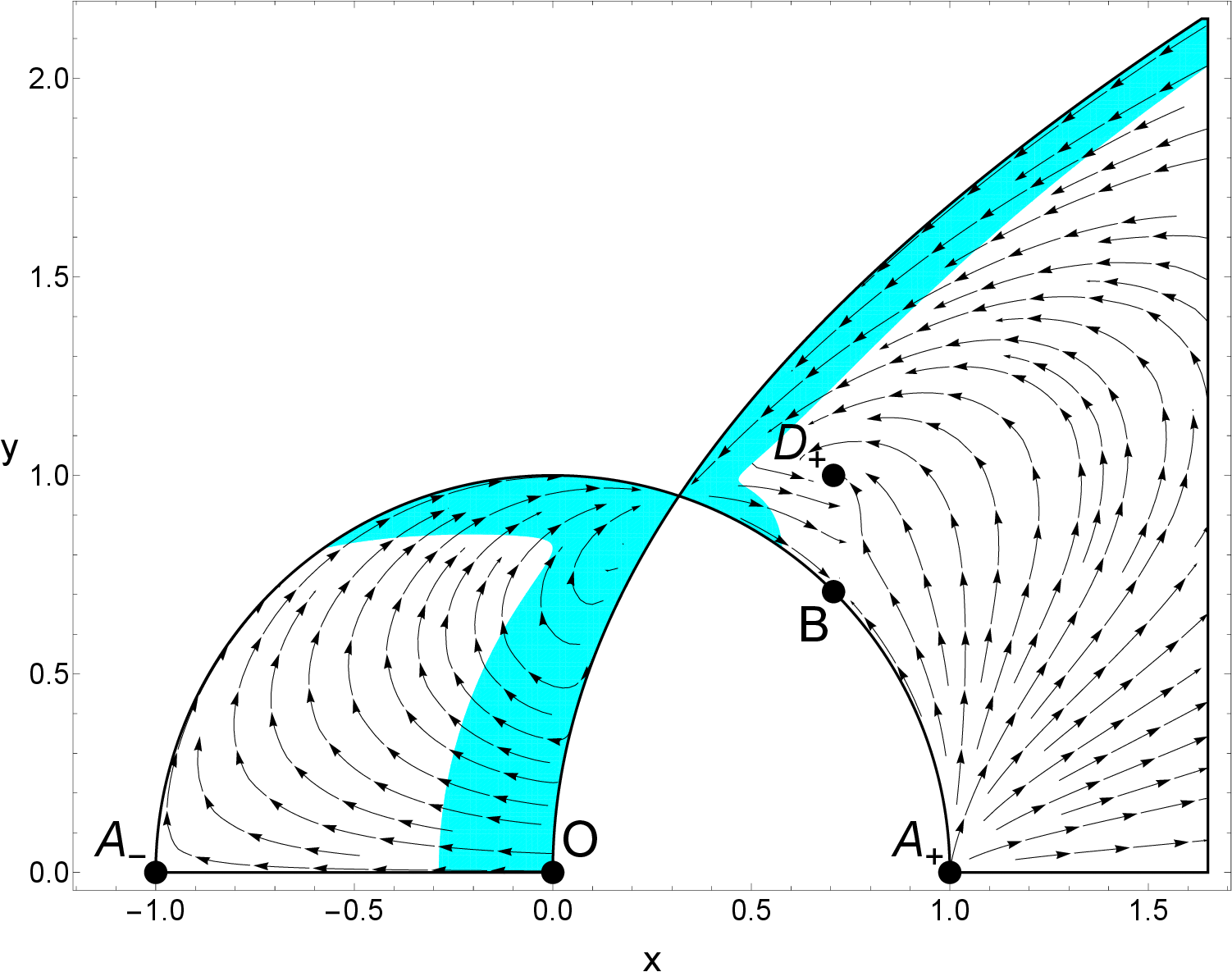}
    \caption{The parameter values are $\lambda=\sqrt{3}$ and $k=\sqrt{8}$. $B$ is an unstable centre and $D_{+}$ is a stable spiral. The shaded region represents the part where the phase space is accelerating.}
    \label{fig:withcentre}
\end{figure}

\paragraph{Case (ii).} Next, we consider the case $\lambda=\sqrt{3}$ and $k=\sqrt{8}$, see Fig.~\ref{fig:withcentre}. In this particular case, Point $D_{-}$ coincides with Point $B$, this is true for all $k$. As in the previous case, Point $C$ does not exist in the physical phase plane. Point $O$, as before, is an unstable node and acts as an early-time attractor. Point $A_{-}$ is a saddle and corresponds to the to other possible early-time attractor. According to Table~\ref{noncontable2}, we note that Point $B$ has eigenvalues $0$ and $-3/2$, which means that we are dealing with a non-hyperbolic point. This point is a centre and one can verify that it is unstable. While this can be shown rigorously, it essentially follows from the fact that trajectories near $B$ move towards the attractor $D_{+}$, which is a stable spiral. It has eigenvalues $-3/4 \pm \mathrm{i} \sqrt{15}/4$. Similar to the previous case, Point $O$ is an early-time attractor corresponding to an early-time universe undergoing accelerated expansion. There is no late-time attractor within the acceleration region.

\paragraph{Case (iii).} We will briefly comment on the case where $\lambda=3/\sqrt{2}$ and $k=2/\sqrt{3}$. For this particular choice, Points $C$ and $D_{-}$ do not exist, while the Point $D_{+}$ is located at the intersection of the two regions of the phase space. This case is mathematically quite interesting, however, less so from a physical point of view. Of mathematical interest are the following facts: $D_{+}$ is a critical point of the system, however, both the numerators and the denominators of~(\ref{noncon1}) and~(\ref{noncon4}) vanish while giving a finite limit. The stability matrix is singular at this point, meaning that linear stability theory cannot be used. We will not discuss this case further.

\begin{figure}[!htb]
    \centering
    \includegraphics[width=0.95\textwidth]{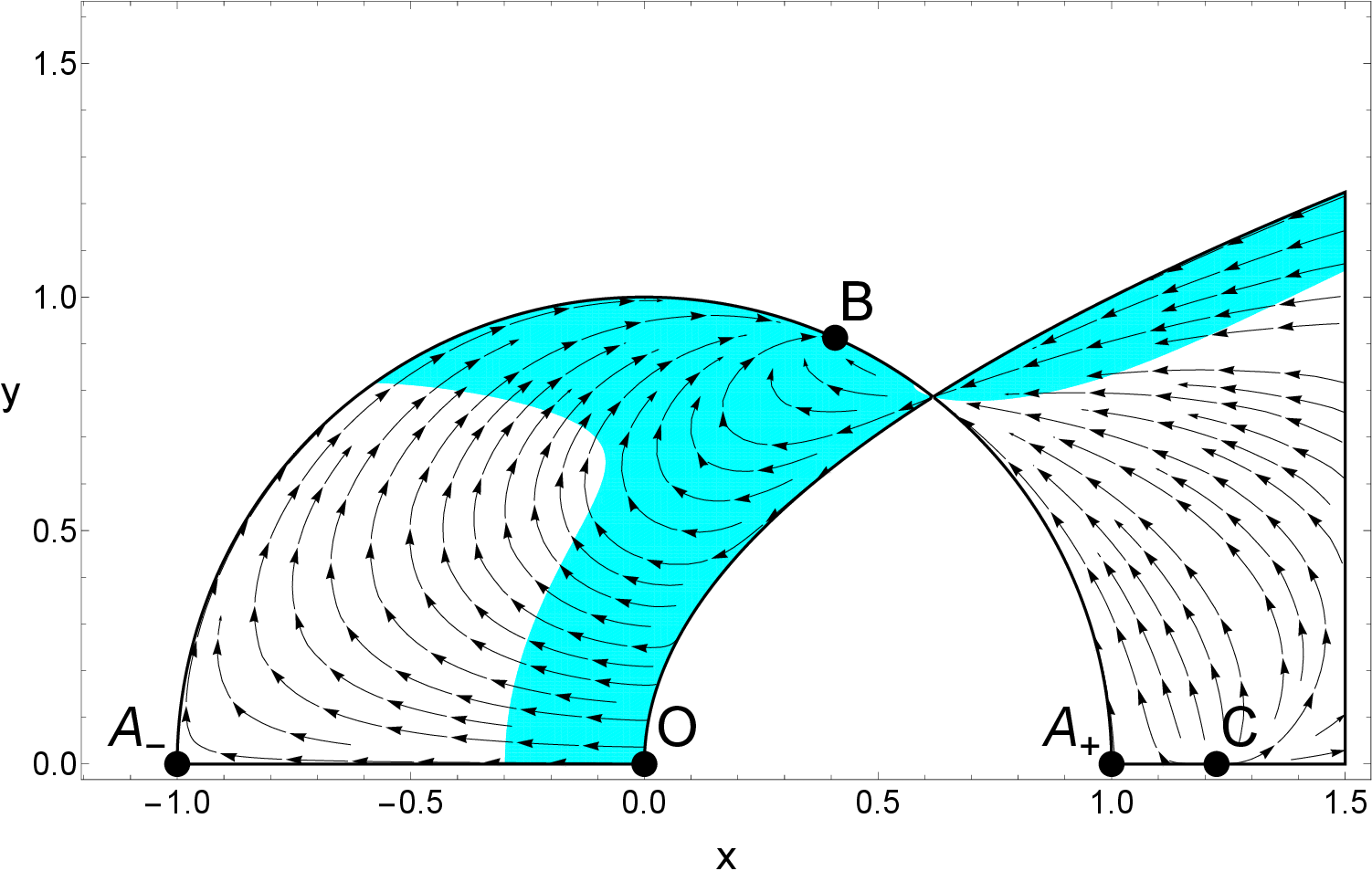}
    \caption{The parameter values are $\lambda=1$ and $k=1$. The shaded region represents the part where the phase space is accelerating.}
    \label{fig:kandlambdaone}
\end{figure}

\paragraph{Case (iv).} Next, we consider the case $\lambda=k=1$, which turns out to be physically interesting as a cosmological model, see Fig.~\ref{fig:kandlambdaone}. There are two unstable nodes, Points $O$ and $C$, which act as early-time attractors. As in the previous cases, Point $O$ corresponds to an early-time universe undergoing accelerated expansion. Trajectories starting near $O$ will eventually leave the acceleration region and be partially attracted to the saddle Point $A_{-}$, after which they will reach the late-time attractor, the stable node at $B$. This point is also in the accelerated region, which means that this model not only allows for early-time acceleration (inflation) but also for late-time accelerated expansion. The effective equation of state parameter at $B$ is $\widetilde{w}=-2/3$, as can be seen from Table~\ref{noncontable}. All trajectories starting out in the right part of the phase space will also be attracted to $B$, making this the global attractor of the system. We remark that, in this sense, the dynamical behaviour is similar to that shown in Fig.~\ref{fig:constfunctregion4}.

\begin{figure}[!htb]
    \centering
    \includegraphics[width=0.95\textwidth]{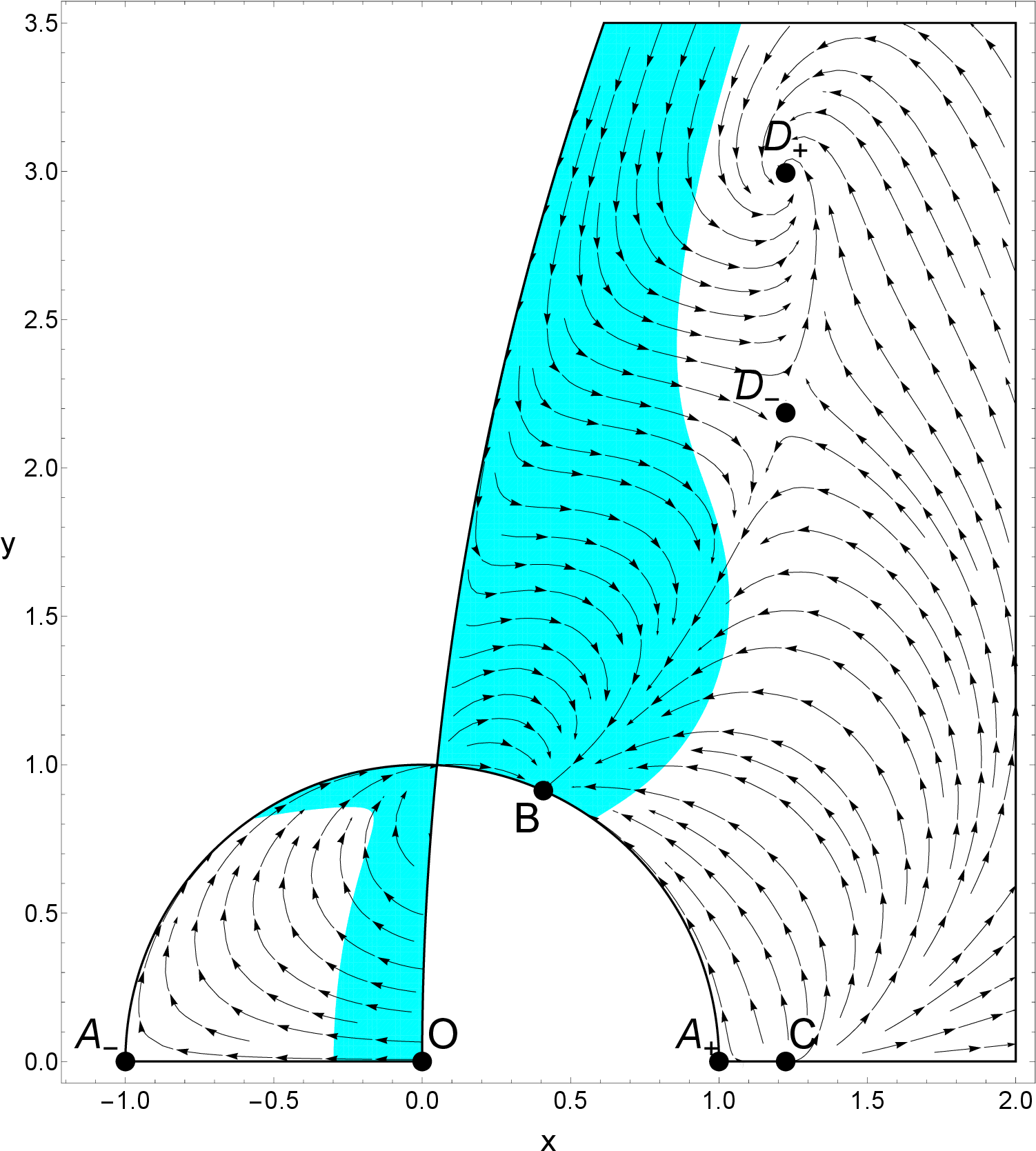}
    \caption{The parameter values are $\lambda=1$ and $k=20$. The eigenvalues of $D_{+}$ are $-0.75 \pm 1.11193 i$ and the eigenvalues of $D_{-}$ are $-2.34057$ and $0.840574$. The shaded area represents the part of the phase space where there is accelerated expansion.}
    \label{fig:ktwenty}
\end{figure}

\paragraph{Case (v).} We now present the case $\lambda=1$ and $k=20$, see Fig.~\ref{fig:ktwenty}. Here all the fixed points we found analytically exist within the physical phase space. This model not only has a rich dynamical structure, but is also of physical relevance. We have two early-time attractors, Point $O$ in the acceleration region, similar to the previous models, and Point $C$. Notice that Point $C$ always satisfies $\widetilde{w} > 1$, and so is not of physical interest. We are therefore most interested in trajectories starting from Point $O$. These will initially move towards $A_{-}$, before leaving the left part of the phase space. By doing so, they will enter the acceleration region and move towards $B$ where $\widetilde{w}=-2/3$. Other than the various complications introduced by the other critical points, and the more complicated phase space structure, the physical situation is again somewhat similar to those shown in Fig.~\ref{fig:constfunctregion4} and Fig.~\ref{fig:kandlambdaone}. Let us also mention that Point $D_{+}$ represents scaling solutions as the effective equation of state parameter is zero, and the universe evolves as if it were only matter dominated while also containing the scalar field.

\begin{figure}[!htb]
    \centering
    \includegraphics[width=0.95\textwidth]{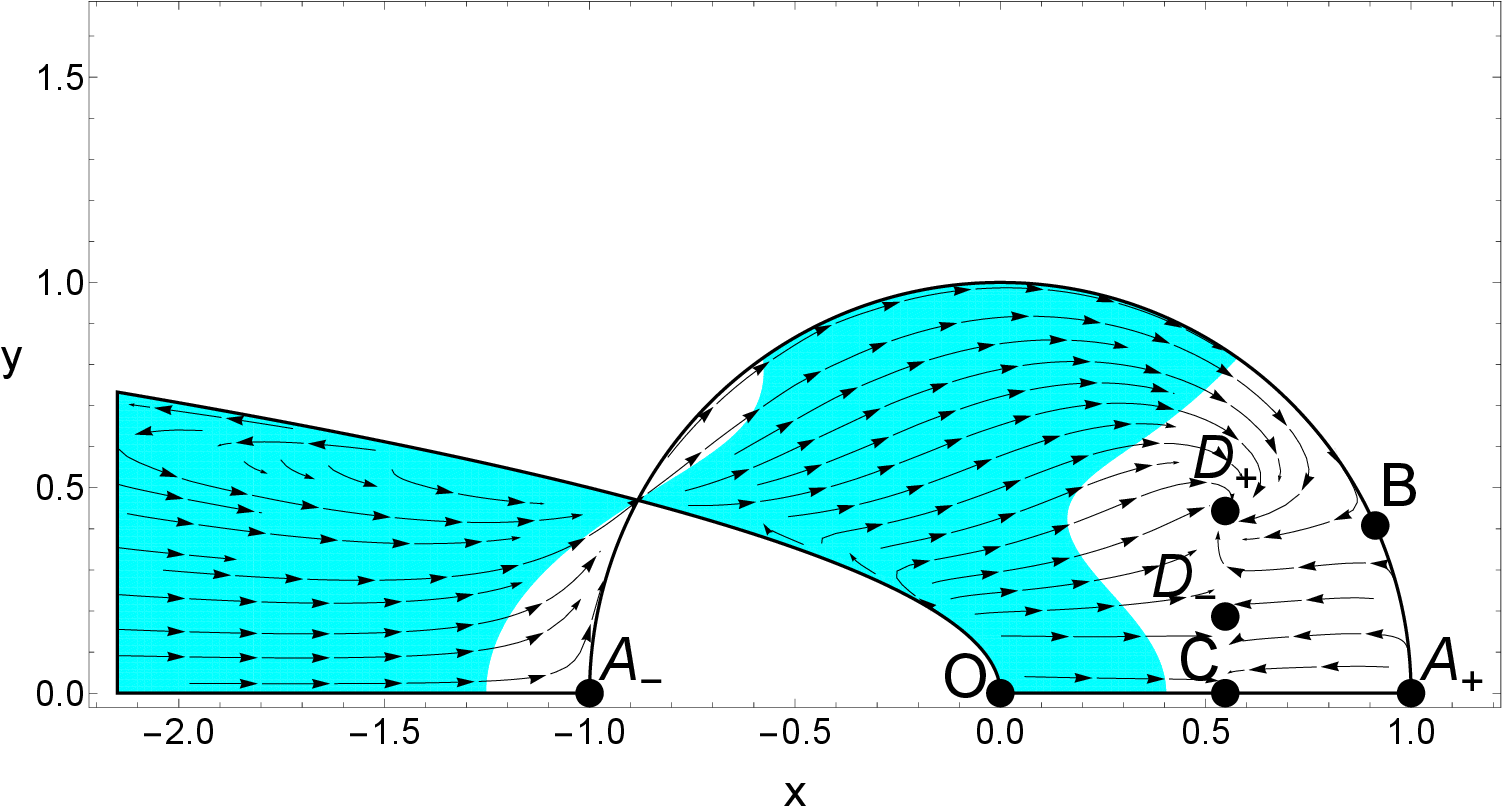}
    \caption{The parameter values are $\lambda=\sqrt{5}$ and $k=-1/4$. The eigenvalues of $D_{+}$ are $-0.75 \pm 0.3654 \mathrm{i}$ and for $D_{-}$ are $-1.6598$ and $0.1598$. The shaded area represents the part of the phase space where there is accelerated expansion.}
    \label{fig:final}
\end{figure}

\paragraph{Case (vi).} To complete this section, we consider a case where the sign of the coupling is negative. We set $k=-1/4<0$ and $\lambda=\sqrt{5}$, and the phase plane is shown in Fig.~\ref{fig:final}. This model displays significantly different features than the cases where the coupling is positive. Point $O$ can still be seen as an early-time attractor in the acceleration region. However, depending on the chosen initial condition, trajectories will either terminate at Point $D_{+}$ or Point $C$. Such trajectories can come close to the saddle Point $D_{-}$, where $\widetilde{w}=1/2$. However, the effective equation of state parameter is also quite large at the other two points, meaning that one cannot have a model with a late-time behaviour close to a matter dominated universe. None of the late-time attractors appear in the acceleration region. Moreover, the left-hand side of the phase plane indicates the presence of a critical point at infinity, which will be the source of the trajectories coming from this left region. While this model displays many interesting mathematical features, it appears to be of more limited physical relevance.

\clearpage

\section{Conclusions and discussions}
\label{sec:disc}

The entire field of cosmology has seen remarkable progress in recent decades, including the use of dynamical system techniques to study the background behaviour of cosmological models. These techniques offer a systematic approach to understanding the underlying dynamics, which allows us to investigate the suitability of such models as realistic approximations of the universe. Our analysis involves mapping the cosmological equations onto a phase space, a step which relies heavily on the choice of suitable variables. This is rather non-trivial as various different variables could be employed and there is no particular reason to prefer one set of variables over any other. We therefore work with those variables that are known to be well suited for our task, see the review~\cite{Bahamonde:2017ize}.

One of the main motivations of this work was to study models derived from a variational principle, in particular, we used Brown's approach for the formulation of the perfect fluid Langrangian for the cosmological matter. This approach allowed us to introduce new coupling terms, including boundary term couplings which have not been studied before in this context using fluids. So far, we have considered an algebraic vector coupling of the form $f(n,s,\phi)B^\mu J_{\mu}$ and noted that, in cosmology, one obtains the highly restrictive condition that $f$ is proportional to $n$. Therefore, such a coupling does not yield interesting phenomena. We therefore focused on a derivative type coupling, $f(n,s,\phi) B^\mu \partial_\mu \phi$, motivated by previous work~\cite{Boehmer:2015sha}. In that previous work, the coupling $f(n,s,\phi) J^\mu \partial_\mu \phi$ led to a minor change of the phase space compared to \cite{Copeland:1997et}: the critical point at the origin moves along the $x$-axis, depending on the choice of parameters. Our new coupling allows for a significantly different dynamical behaviour with features unseen before. Of particular interest to cosmology are situations where the model evolves through two periods of accelerated expansion, which replicate inflation and dark energy in a single model.

To gain an initial understanding of the resulting cosmological model, we began by studying the constant interaction model. This displayed similar behaviour to the well-known exponential potential quintessence model~\cite{Copeland:1997et}, where no interacting term is present. In fact, through a carefully chosen change of variables, we were able to arrive at a phase space which mirrors the one studied by those authors. These results demonstrate that our model can be seen as a natural extension of previous work.

We then proceeded to consider an interaction of the form $n^{\alpha/(2(1+w))}V^{-\alpha/2}$. This choice was motivated by the fact that couplings of this form will not increase the dimension of the phase space: it will remain two-dimensional. The key advantage of this assumption is that one can directly compare results with many previously studied models. In particular, we focused on the matter dominated case, $w=0$, and chose $\alpha=2$. This choice leads to a rich dynamical structure with several distinct scenarios that can be of physical relevance. We were able to obtain solutions with early-time inflationary attractors, as well as late-time acceleration. Our models also included scaling solutions, which have received recent attention, see~\cite{Copeland:2023zqz}, as they may help to resolve the \emph{Hubble tension}, that is, the discrepancy between the value of the Hubble constant inferred from measurements of the early universe and those derived from more recent observations~\cite{DiValentino:2021izs,Yang:2021hxg}.

Our approach to constructing coupled models lends itself to a significant amount of further study. First of all, one can study the constant coupling model in the radiation dominated universe. Our preliminary work suggests that the results are qualitatively similar to the matter dominated case, which is why we did not include them here, for the constant coupling model. One could attempt to present a comprehensive study for all $w$, however, this would not be without challenge as the convoluted equations would make analysing the stability at fixed points difficult. 

Regarding the non-constant coupling model we proposed, there are three obvious extensions to our work, namely the cases $\alpha\in\{-2,-1,1\}$. For integer values of $\alpha$, the phase will remain two-dimensional. However, one encounters other challenges which can be seen in equation~(\ref{fried-non1}). For example, when setting $\alpha=-2$ one should eliminate $y$ from the equation instead of the matter variable $\sigma$. This is unusual and has rarely been considered in the past. As a starting point, one would have to go back to the baseline model~\cite{Copeland:1997et} and study it using a different choice of variables.  In this way, comparisons could be drawn. For large values of $\alpha$, on the other hand, one is dealing with a polynomial of a high degree, which is difficult to handle. In such cases, the best way forward would be to eliminate the variable $x$. This is equally unusual, and has also not been considered in the past. Given the complexities of these models, we are not able to predict the qualitative features of the resulting systems. The shape of the phase space alone changes significantly when varying the parameter $\alpha$.

On top of all of this, one could, of course, drop the assumption of working with an exponential potential. It would be interesting to study our coupled models for power-law potentials or others. Most of what has been done here will have to be re-investigated from scratch. For example, it is not even clear which types of couplings will admit a two-dimensional phase space. One would expect the constant coupling models to be similar to the uncoupled models, however, we refrain from speculating about results beyond this most basic of statements.

\subsection*{Acknowledgements}

AD would like to thank David Sheard for useful discussions. Antonio d'Alfonso del Sordo is supported by the Engineering and Physical Sciences Research Council EP/R513143/1 \& EP/T517793/1. 

\appendix
\section{Fixed points and classification}\label{sec:AppendixA}
For a continuously differentiable $\mathbf{f}=(f_1,\ldots,f_n)$, depending on the variable $\mathbf{x}=(x_1,\ldots,x_n)$, the initial value problem 
\begin{equation}
    \dv{\mathbf{x}}{t}=\mathbf{f}(\mathbf{x}), \quad\mathbf{x}(0)=\mathbf{x}_0, \label{eqn:dynsys}
\end{equation}
admits a unique solution, that is, the solution $\mathbf{x}(t)$ with the given initial condition $\mathbf{x}_0$ at $t=0$. 
\emph{Fixed (or critical) points} are points $\mathbf{x}^{*}$ such that $\mathbf{f}(\mathbf{x}^{*})=\mathbf{0}$. These points are also known as \emph{steady states} or \emph{equilibria} of the system.

To determine the behaviour of trajectories near those fixed points, we can linearise the system around its critical point, by using a Taylor expansion for $\mathbf{f}$ in the neighbourhood of the fixed point. The dynamics of the linearised system are qualitatively equivalent to the original system. The eigenvalues of the matrix $\boldsymbol{\nabla}\mathbf{f}(\mathbf{x}^*)$, known as the \emph{Jacobian} matrix or \emph{stability} matrix, contain the information about the local behaviour of $\mathbf{f}$ near $\mathbf{x}^*$. One generally speaks of stability or instability:
\begin{itemize}
    \item if all eigenvalues have positive real parts, we have an unstable fixed point or repeller
    \item if all eigenvalues have negative real parts, we have a stable fixed point or attractor
    \item if at least two eigenvalues have real parts with opposite signs, the corresponding fixed point is called a saddle point
    \item if an eigenvalue is zero and at least one other eigenvalue has positive real parts, we have an unstable point
    \item if an eigenvalue is zero and all other eigenvalues have negative real parts, linear stability theory does not suffice.
\end{itemize}
For more details, see for example~\cite{B_hmer_2016}.

\addcontentsline{toc}{section}{References}
\bibliographystyle{jhepmodstyle}
\bibliography{bib}

\providecommand{\href}[2]{#2}\begingroup\raggedright\begin{thebibliography}{10}

\bibitem{LIGOScientific:2016aoc}
{\bf LIGO Scientific, Virgo} Collaboration, B.~P. Abbott et~al., {\it
  {Observation of Gravitational Waves from a Binary Black Hole Merger}},  Phys.
  Rev. Lett. {\bf 116} (2016), no.~6 061102,
  [\href{http://arxiv.org/abs/1602.03837}{{\tt arXiv:1602.03837}}].

\bibitem{Planck:2018vyg}
{\bf Planck} Collaboration, N.~Aghanim et~al., {\it {Planck 2018 results. VI.
  Cosmological parameters}},  Astron. Astrophys. {\bf 641} (2020) A6,
  [\href{http://arxiv.org/abs/1807.06209}{{\tt arXiv:1807.06209}}]. [Erratum:
  Astron.Astrophys. 652, C4 (2021)].

\bibitem{Will:2018bme}
C.~M. Will, {\em {Theory and Experiment in Gravitational Physics}}.
\newblock Cambridge University Press, 9, 2018.

\bibitem{Ishak:2018his}
M.~Ishak, {\it {Testing General Relativity in Cosmology}},  Living Rev. Rel.
  {\bf 22} (2019), no.~1 1, [\href{http://arxiv.org/abs/1806.10122}{{\tt
  arXiv:1806.10122}}].

\bibitem{SupernovaSearchTeam:1998fmf}
{\bf Supernova Search Team} Collaboration, A.~G. Riess et~al., {\it
  {Observational evidence from supernovae for an accelerating universe and a
  cosmological constant}},  Astron. J. {\bf 116} (1998) 1009--1038,
  [\href{http://arxiv.org/abs/astro-ph/9805201}{{\tt astro-ph/9805201}}].

\bibitem{SupernovaCosmologyProject:1998vns}
{\bf Supernova Cosmology Project} Collaboration, S.~Perlmutter et~al., {\it
  {Measurements of $\Omega$ and $\Lambda$ from 42 High Redshift Supernovae}},
  Astrophys. J. {\bf 517} (1999) 565--586,
  [\href{http://arxiv.org/abs/astro-ph/9812133}{{\tt astro-ph/9812133}}].

\bibitem{Copeland:2006wr}
E.~J. Copeland, M.~Sami, and S.~Tsujikawa, {\it {Dynamics of dark energy}},
  Int. J. Mod. Phys. D {\bf 15} (2006) 1753--1936,
  [\href{http://arxiv.org/abs/hep-th/0603057}{{\tt hep-th/0603057}}].

\bibitem{Einstein:1917ce}
A.~Einstein, {\it {Cosmological Considerations in the General Theory of
  Relativity}},  Sitzungsber. Preuss. Akad. Wiss. Berlin (Math. Phys.) (1917)
  142--152.

\bibitem{Weinberg:1988cp}
S.~Weinberg, {\it {The Cosmological Constant Problem}},  Rev. Mod. Phys. {\bf
  61} (1989) 1--23.

\bibitem{Zlatev:1998tr}
I.~Zlatev, L.-M. Wang, and P.~J. Steinhardt, {\it {Quintessence, cosmic
  coincidence, and the cosmological constant}},  Phys. Rev. Lett. {\bf 82}
  (1999) 896--899, [\href{http://arxiv.org/abs/astro-ph/9807002}{{\tt
  astro-ph/9807002}}].

\bibitem{Sadjadi:2006qp}
H.~M. Sadjadi and M.~Alimohammadi, {\it {Cosmological coincidence problem in
  interactive dark energy models}},  Phys. Rev. D {\bf 74} (2006) 103007,
  [\href{http://arxiv.org/abs/gr-qc/0610080}{{\tt gr-qc/0610080}}].

\bibitem{Tsujikawa:2013fta}
S.~Tsujikawa, {\it {Quintessence: A Review}},  Class. Quant. Grav. {\bf 30}
  (2013) 214003, [\href{http://arxiv.org/abs/1304.1961}{{\tt
  arXiv:1304.1961}}].

\bibitem{Tamanini:2014mpa}
N.~Tamanini, {\it {Dynamics of cosmological scalar fields}},  Phys. Rev. D {\bf
  89} (2014) 083521, [\href{http://arxiv.org/abs/1401.6339}{{\tt
  arXiv:1401.6339}}].

\bibitem{Bohmer:2016ome}
C.~G. B{\"o}hmer, {\em {Introduction to General Relativity and Cosmology}}.
\newblock Essential Textbooks in Physics. World Scientific, 12, 2016.

\bibitem{Urena-Lopez:2020npg}
L.~A. Ure\~na L\'opez and N.~Roy, {\it {Generalized tracker quintessence models
  for dark energy}},  Phys. Rev. D {\bf 102} (2020), no.~6 063510,
  [\href{http://arxiv.org/abs/2007.08873}{{\tt arXiv:2007.08873}}].

\bibitem{Magana:2012ph}
J.~Magana and T.~Matos, {\it {A brief Review of the Scalar Field Dark Matter
  model}},  J. Phys. Conf. Ser. {\bf 378} (2012) 012012,
  [\href{http://arxiv.org/abs/1201.6107}{{\tt arXiv:1201.6107}}].

\bibitem{CANTATA:2021ktz}
{\bf CANTATA} Collaboration, E.~N. Saridakis, R.~Lazkoz, V.~Salzano,
  P.~Vargas~Moniz, S.~Capozziello, J.~Beltr\'an~Jim\'enez, M.~De~Laurentis, and
  G.~J. Olmo, eds., {\em {Modified Gravity and Cosmology}: {An Update by the
  CANTATA Network}}.
\newblock Springer, 2021.

\bibitem{Koyama:2018som}
K.~Koyama, {\it {Gravity beyond general relativity}},  Int. J. Mod. Phys. D
  {\bf 27} (2018), no.~15 1848001.

\bibitem{Joyce:2014kja}
A.~Joyce, B.~Jain, J.~Khoury, and M.~Trodden, {\it {Beyond the Cosmological
  Standard Model}},  Phys. Rept. {\bf 568} (2015) 1--98,
  [\href{http://arxiv.org/abs/1407.0059}{{\tt arXiv:1407.0059}}].

\bibitem{Billyard:2000bh}
A.~P. Billyard and A.~A. Coley, {\it {Interactions in scalar field cosmology}},
   Phys. Rev. D {\bf 61} (2000) 083503,
  [\href{http://arxiv.org/abs/astro-ph/9908224}{{\tt astro-ph/9908224}}].

\bibitem{Farrar:2003uw}
G.~R. Farrar and P.~J.~E. Peebles, {\it {Interacting dark matter and dark
  energy}},  Astrophys. J. {\bf 604} (2004) 1--11,
  [\href{http://arxiv.org/abs/astro-ph/0307316}{{\tt astro-ph/0307316}}].

\bibitem{Guo:2007zk}
Z.-K. Guo, N.~Ohta, and S.~Tsujikawa, {\it {Probing the Coupling between Dark
  Components of the Universe}},  Phys. Rev. D {\bf 76} (2007) 023508,
  [\href{http://arxiv.org/abs/astro-ph/0702015}{{\tt astro-ph/0702015}}].

\bibitem{Caldera-Cabral:2008yyo}
G.~Caldera-Cabral, R.~Maartens, and L.~A. Urena-Lopez, {\it {Dynamics of
  interacting dark energy}},  Phys. Rev. D {\bf 79} (2009) 063518,
  [\href{http://arxiv.org/abs/0812.1827}{{\tt arXiv:0812.1827}}].

\bibitem{He:2008tn}
J.-H. He and B.~Wang, {\it {Effects of the interaction between dark energy and
  dark matter on cosmological parameters}},  JCAP {\bf 06} (2008) 010,
  [\href{http://arxiv.org/abs/0801.4233}{{\tt arXiv:0801.4233}}].

\bibitem{Pereira:2008at}
S.~H. Pereira and J.~F. Jesus, {\it {Can Dark Matter Decay in Dark Energy?}},
  Phys. Rev. D {\bf 79} (2009) 043517,
  [\href{http://arxiv.org/abs/0811.0099}{{\tt arXiv:0811.0099}}].

\bibitem{Valiviita:2009nu}
J.~Valiviita, R.~Maartens, and E.~Majerotto, {\it {Observational constraints on
  an interacting dark energy model}},  Mon. Not. Roy. Astron. Soc. {\bf 402}
  (2010) 2355--2368, [\href{http://arxiv.org/abs/0907.4987}{{\tt
  arXiv:0907.4987}}].

\bibitem{Bamba:2012cp}
K.~Bamba, S.~Capozziello, S.~Nojiri, and S.~D. Odintsov, {\it {Dark energy
  cosmology: the equivalent description via different theoretical models and
  cosmography tests}},  Astrophys. Space Sci. {\bf 342} (2012) 155--228,
  [\href{http://arxiv.org/abs/1205.3421}{{\tt arXiv:1205.3421}}].

\bibitem{Joyce:2016}
A.~Joyce, L.~Lombriser, and F.~Schmidt, {\it {Dark Energy Versus Modified
  Gravity}},  Ann. Rev. Nucl. Part. Sci. {\bf 66} (2016) 95--122,
  [\href{http://arxiv.org/abs/1601.06133}{{\tt arXiv:1601.06133}}].

\bibitem{Magnano:1987zz}
G.~Magnano, M.~Ferraris, and M.~Francaviglia, {\it {Nonlinear gravitational
  Lagrangians}},  Gen. Rel. Grav. {\bf 19} (1987) 465.

\bibitem{Capozziello:2007ec}
S.~Capozziello and M.~Francaviglia, {\it {Extended Theories of Gravity and
  their Cosmological and Astrophysical Applications}},  Gen. Rel. Grav. {\bf
  40} (2008) 357--420, [\href{http://arxiv.org/abs/0706.1146}{{\tt
  arXiv:0706.1146}}].

\bibitem{Sotiriou:2008rp}
T.~P. Sotiriou and V.~Faraoni, {\it {f(R) Theories Of Gravity}},  Rev. Mod.
  Phys. {\bf 82} (2010) 451--497, [\href{http://arxiv.org/abs/0805.1726}{{\tt
  arXiv:0805.1726}}].

\bibitem{DeFelice:2010aj}
A.~De~Felice and S.~Tsujikawa, {\it {f(R) theories}},  Living Rev. Rel. {\bf
  13} (2010) 3, [\href{http://arxiv.org/abs/1002.4928}{{\tt arXiv:1002.4928}}].

\bibitem{Capozziello:2002rd}
S.~Capozziello, {\it {Curvature quintessence}},  Int. J. Mod. Phys. D {\bf 11}
  (2002) 483--492, [\href{http://arxiv.org/abs/gr-qc/0201033}{{\tt
  gr-qc/0201033}}].

\bibitem{Ferraro:2006jd}
R.~Ferraro and F.~Fiorini, {\it {Modified teleparallel gravity: Inflation
  without inflaton}},  Phys. Rev. D {\bf 75} (2007) 084031,
  [\href{http://arxiv.org/abs/gr-qc/0610067}{{\tt gr-qc/0610067}}].

\bibitem{Nojiri:2006je}
S.~Nojiri, S.~D. Odintsov, and M.~Sami, {\it {Dark energy cosmology from
  higher-order, string-inspired gravity and its reconstruction}},  Phys. Rev. D
  {\bf 74} (2006) 046004, [\href{http://arxiv.org/abs/hep-th/0605039}{{\tt
  hep-th/0605039}}].

\bibitem{Nojiri:2010wj}
S.~Nojiri and S.~D. Odintsov, {\it {Unified cosmic history in modified gravity:
  from F(R) theory to Lorentz non-invariant models}},  Phys. Rept. {\bf 505}
  (2011) 59--144, [\href{http://arxiv.org/abs/1011.0544}{{\tt
  arXiv:1011.0544}}].

\bibitem{Harko:2011kv}
T.~Harko, F.~S.~N. Lobo, S.~Nojiri, and S.~D. Odintsov, {\it {$f(R,T)$
  gravity}},  Phys. Rev. D {\bf 84} (2011) 024020,
  [\href{http://arxiv.org/abs/1104.2669}{{\tt arXiv:1104.2669}}].

\bibitem{Clifton:2011jh}
T.~Clifton, P.~G. Ferreira, A.~Padilla, and C.~Skordis, {\it {Modified Gravity
  and Cosmology}},  Phys. Rept. {\bf 513} (2012) 1--189,
  [\href{http://arxiv.org/abs/1106.2476}{{\tt arXiv:1106.2476}}].

\bibitem{Granda:2014zea}
L.~N. Granda and D.~F. Jimenez, {\it {Dark Energy from Gauss-Bonnet and
  non-minimal couplings}},  Phys. Rev. D {\bf 90} (2014), no.~12 123512,
  [\href{http://arxiv.org/abs/1411.4203}{{\tt arXiv:1411.4203}}].

\bibitem{Bahamonde:2015hza}
S.~Bahamonde and M.~Wright, {\it {Teleparallel quintessence with a nonminimal
  coupling to a boundary term}},  Phys. Rev. D {\bf 92} (2015), no.~8 084034,
  [\href{http://arxiv.org/abs/1508.06580}{{\tt arXiv:1508.06580}}]. [Erratum:
  Phys.Rev.D 93, 109901 (2016)].

\bibitem{Nojiri:2017ncd}
S.~Nojiri, S.~D. Odintsov, and V.~K. Oikonomou, {\it {Modified Gravity Theories
  on a Nutshell: Inflation, Bounce and Late-time Evolution}},  Phys. Rept. {\bf
  692} (2017) 1--104, [\href{http://arxiv.org/abs/1705.11098}{{\tt
  arXiv:1705.11098}}].

\bibitem{Boehmer:2021aji}
C.~G. B{\"o}hmer and E.~Jensko, {\it {Modified gravity: A unified approach}},
  Phys. Rev. D {\bf 104} (2021), no.~2 024010,
  [\href{http://arxiv.org/abs/2103.15906}{{\tt arXiv:2103.15906}}].

\bibitem{Boehmer:2015kta}
C.~G. B\"ohmer, N.~Tamanini, and M.~Wright, {\it {Interacting quintessence from
  a variational approach Part I: algebraic couplings}},  Phys. Rev. D {\bf 91}
  (2015), no.~12 123002, [\href{http://arxiv.org/abs/1501.06540}{{\tt
  arXiv:1501.06540}}].

\bibitem{Boehmer:2015sha}
C.~G. B\"ohmer, N.~Tamanini, and M.~Wright, {\it {Interacting quintessence from
  a variational approach Part II: derivative couplings}},  Phys. Rev. D {\bf
  91} (2015), no.~12 123003, [\href{http://arxiv.org/abs/1502.04030}{{\tt
  arXiv:1502.04030}}].

\bibitem{Boehmer:2008av}
C.~G. B{\"o}hmer, G.~Caldera-Cabral, R.~Lazkoz, and R.~Maartens, {\it Dynamics
  of dark energy with a coupling to dark matter},  Phys. Rev. D {\bf 78} (2008)
  023505, [\href{http://arxiv.org/abs/0801.1565}{{\tt arXiv:0801.1565}}].

\bibitem{Tamanini:2015iia}
N.~Tamanini, {\it {Phenomenological models of dark energy interacting with dark
  matter}},  Phys. Rev. D {\bf 92} (2015), no.~4 043524,
  [\href{http://arxiv.org/abs/1504.07397}{{\tt arXiv:1504.07397}}].

\bibitem{Brown:1992kc}
J.~D. Brown, {\it {Action functionals for relativistic perfect fluids}},
  Class. Quant. Grav. {\bf 10} (1993) 1579--1606,
  [\href{http://arxiv.org/abs/gr-qc/9304026}{{\tt gr-qc/9304026}}].

\bibitem{Koivisto:2015qua}
T.~S. Koivisto, E.~N. Saridakis, and N.~Tamanini, {\it {Scalar-Fluid theories:
  cosmological perturbations and large-scale structure}},  JCAP {\bf 09} (2015)
  047, [\href{http://arxiv.org/abs/1505.07556}{{\tt arXiv:1505.07556}}].

\bibitem{Kadam:2023ufk}
S.~A. Kadam, N.~P. Thakkar, and B.~Mishra, {\it {Dynamical system analysis in
  teleparallel gravity with boundary term}},  Eur. Phys. J. C {\bf 83} (2023),
  no.~9 809, [\href{http://arxiv.org/abs/2306.06677}{{\tt arXiv:2306.06677}}].

\bibitem{Bahamonde:2017ize}
S.~Bahamonde, C.~G. B\"ohmer, S.~Carloni, E.~J. Copeland, W.~Fang, and
  N.~Tamanini, {\it {Dynamical systems applied to cosmology: dark energy and
  modified gravity}},  Phys. Rept. {\bf 775-777} (2018) 1--122,
  [\href{http://arxiv.org/abs/1712.03107}{{\tt arXiv:1712.03107}}].

\bibitem{Boehmer:2022wln}
C.~G. B{\"o}hmer, E.~Jensko, and R.~Lazkoz, {\it {Cosmological dynamical
  systems in modified gravity}},  Eur. Phys. J. C {\bf 82} (2022), no.~6 500,
  [\href{http://arxiv.org/abs/2201.09588}{{\tt arXiv:2201.09588}}].

\bibitem{Dutta:2017fjw}
J.~Dutta, W.~Khyllep, E.~N. Saridakis, N.~Tamanini, and S.~Vagnozzi, {\it
  {Cosmological dynamics of mimetic gravity}},  JCAP {\bf 02} (2018) 041,
  [\href{http://arxiv.org/abs/1711.07290}{{\tt arXiv:1711.07290}}].

\bibitem{Khyllep:2021pcu}
W.~Khyllep, A.~Paliathanasis, and J.~Dutta, {\it {Cosmological solutions and
  growth index of matter perturbations in $f(Q)$ gravity}},  Phys. Rev. D {\bf
  103} (2021), no.~10 103521, [\href{http://arxiv.org/abs/2103.08372}{{\tt
  arXiv:2103.08372}}].

\bibitem{Copeland:1997et}
E.~J. Copeland, A.~R. Liddle, and D.~Wands, {\it {Exponential potentials and
  cosmological scaling solutions}},  Phys. Rev. D {\bf 57} (1998) 4686--4690,
  [\href{http://arxiv.org/abs/gr-qc/9711068}{{\tt gr-qc/9711068}}].

\bibitem{Boehmer:2023fyl}
C.~G. B{\"o}hmer and E.~Jensko, {\it {Modified gravity: A unified approach to
  metric-affine models}},  J. Math. Phys. {\bf 64} (2023), no.~8 082505,
  [\href{http://arxiv.org/abs/2301.11051}{{\tt arXiv:2301.11051}}].

\bibitem{Saadeh:2016sak}
D.~Saadeh, S.~M. Feeney, A.~Pontzen, H.~V. Peiris, and J.~D. McEwen, {\it {How
  isotropic is the Universe?}},  Phys. Rev. Lett. {\bf 117} (2016), no.~13
  131302, [\href{http://arxiv.org/abs/1605.07178}{{\tt arXiv:1605.07178}}].

\bibitem{Efstathiou:2020wem}
G.~Efstathiou and S.~Gratton, {\it {The evidence for a spatially flat
  Universe}},  Mon. Not. Roy. Astron. Soc. {\bf 496} (2020), no.~1 L91--L95,
  [\href{http://arxiv.org/abs/2002.06892}{{\tt arXiv:2002.06892}}].

\bibitem{Boehmer:2011tp}
C.~G. B{\"o}hmer, N.~Chan, and R.~Lazkoz, {\it {Dynamics of dark energy models
  and centre manifolds}},  Phys. Lett. B {\bf 714} (2012) 11--17,
  [\href{http://arxiv.org/abs/1111.6247}{{\tt arXiv:1111.6247}}].

\bibitem{B_hmer_2016}
C.~G. Böhmer and N.~Chan, {\it Dynamical systems in cosmology},  in {\em
  Dynamical and Complex Systems}, pp.~121--156.
\newblock {World} {Scientific} ({Europe}), 12, 2016.

\bibitem{Amendola:1999qq}
L.~Amendola, {\it {Scaling solutions in general nonminimal coupling theories}},
   Phys. Rev. D {\bf 60} (1999) 043501,
  [\href{http://arxiv.org/abs/astro-ph/9904120}{{\tt astro-ph/9904120}}].

\bibitem{Nunes:2000ka}
A.~Nunes, J.~P. Mimoso, and T.~C. Charters, {\it {Scaling solutions from
  interacting fluids}},  Phys. Rev. D {\bf 63} (2001) 083506,
  [\href{http://arxiv.org/abs/gr-qc/0011073}{{\tt gr-qc/0011073}}].

\bibitem{Teixeira:2019tfi}
E.~M. Teixeira, A.~Nunes, and N.~J. Nunes, {\it {Conformally Coupled Tachyonic
  Dark Energy}},  Phys. Rev. D {\bf 100} (2019), no.~4 043539,
  [\href{http://arxiv.org/abs/1903.06028}{{\tt arXiv:1903.06028}}].

\bibitem{Copeland:2023zqz}
E.~J. Copeland, A.~Moss, S.~Sevillano Mu\~noz, and J.~M.~M. White, {\it
  {Scaling solutions as Early Dark Energy resolutions to the Hubble tension}},
  \href{http://arxiv.org/abs/2309.15295}{{\tt arXiv:2309.15295}}.

\bibitem{DiValentino:2021izs}
E.~Di~Valentino, O.~Mena, S.~Pan, L.~Visinelli, W.~Yang, A.~Melchiorri, D.~F.
  Mota, A.~G. Riess, and J.~Silk, {\it {In the realm of the Hubble
  tension\textemdash{}a review of solutions}},  Class. Quant. Grav. {\bf 38}
  (2021), no.~15 153001, [\href{http://arxiv.org/abs/2103.01183}{{\tt
  arXiv:2103.01183}}].

\bibitem{Yang:2021hxg}
W.~Yang, S.~Pan, E.~Di~Valentino, O.~Mena, and A.~Melchiorri, {\it {2021-H0
  odyssey: closed, phantom and interacting dark energy cosmologies}},  JCAP
  {\bf 10} (2021) 008, [\href{http://arxiv.org/abs/2101.03129}{{\tt
  arXiv:2101.03129}}].

\end{thebibliography}\endgroup

\end{document}